\documentclass[lettersize,journal]{IEEEtran}
\usepackage{amsmath,amsfonts}
\usepackage[linesnumbered,ruled]{algorithm2e}
\usepackage{array}
\usepackage[caption=false,font=footnotesize,labelfont=rm,textfont=rm]{subfig}
\usepackage[utf8]{inputenc}
\usepackage{textcomp}
\usepackage{amsmath}
\usepackage{stfloats}
\usepackage{url}
\usepackage{bm}
\usepackage{verbatim}
\usepackage{graphicx}
\usepackage{subfig}
\usepackage{float}
\usepackage{cite}
\usepackage[T1]{fontenc}
\usepackage{amsmath}
\usepackage{setspace}
\usepackage{indentfirst}
\usepackage{color}
\usepackage{xcolor}

\allowdisplaybreaks[4]

\newtheorem{lemma}{\hspace{1em}Lemma}

\def\BibTeX{{\rm B\kern-.05em{\sc i\kern-.025em b}\kern-.08em
   T\kern-.1667em\lower.7ex\hbox{E}\kern-.125emX}}

\begin{document}
\title{Distributed Computation Offloading for Energy Provision Minimization in WP-MEC Networks with Multiple HAPs}

\author{	
	Xiaoying~Liu,~\IEEEmembership{Senior Member,~IEEE},
	Anping~Chen,
	Kechen~Zheng,~\IEEEmembership{Senior Member,~IEEE},
	
	Kaikai~Chi,~\IEEEmembership{Senior Member,~IEEE},
    Bin~Yang,
    and Tarik Taleb,~\IEEEmembership{Senior Member,~IEEE}

	\thanks{X. Liu, A. Chen, K. Zheng, and K. Chi are with the School of Computer Science and Technology, Zhejiang University of Technology, Hangzhou 310023, China. E-mail: \{xiaoyingliu, 221122120218, kechenzheng, kkchi\}@zjut.edu.cn.}
    \thanks{B. Yang is with the School of Computer and Information Engineering, Chuzhou University, Chuzhou 239000, China. E-mail: yangbinchi@gmail.com.}
    \thanks{T. Taleb is with the Faculty of Electrical Engineering and Information
	Technology, Ruhr University Bochum, Bochum 44801, Germany. E-mail:
	tarik.taleb@rub.de.}
}

\maketitle
\begin{abstract}
	This paper investigates a wireless powered mobile edge computing (WP-MEC) network with multiple hybrid access points (HAPs) in a dynamic environment, where wireless devices (WDs) harvest energy from radio frequency (RF) signals of HAPs, and then compute their computation data locally (i.e., local computing mode) or offload it to the chosen HAPs (i.e., edge computing mode). In order to pursue a green computing design, we formulate an optimization problem that minimizes the long-term energy provision of the WP-MEC network subject to the energy, computing delay and computation data demand constraints. The transmit power of HAPs, the duration of the wireless power transfer (WPT) phase, the offloading decisions of WDs, the time allocation for offloading and the CPU frequency for local computing are jointly optimized adapting to the time-varying generated computation data and wireless channels of WDs. To efficiently address the formulated non-convex mixed integer programming (MIP) problem in a distributed manner, we propose a \underline{T}wo-stage \underline{M}ulti-\underline{A}gent deep reinforcement learning-based \underline{D}istributed computation \underline{O}ffloading (TMADO) framework, which consists of a high-level agent and multiple low-level agents. The high-level agent residing in all HAPs optimizes the transmit power of HAPs and the duration of the WPT phase, while each low-level agent residing in each WD optimizes its offloading decision, time allocation for offloading and CPU frequency for local computing. Simulation results show the superiority of the proposed TMADO framework in terms of the energy provision minimization.
\end{abstract}
	 

\section{Introduction}
 
The fast development of Internet of Things (IoT) has driven various new applications, such as automatic navigation and autonomous driving~{\cite{KHAN2019219}}. These new applications have imposed a great demand on the computing capabilities of wireless devices (WDs) since they have computation-intensive and latency-sensitive tasks to be executed~\cite{WXJ2023WPMECSurvey}. However, most WDs in IoT have low computing capabilities. Mobile edge computing (MEC)~\cite{WKL2023JSAC}  has been identified as one of the promising technologies to improve the computing capabilities of WDs, through offloading computing tasks from WDs to surrounding MEC servers acted by access points (APs) or base stations (BSs). In this way, MEC servers' computing capabilities could be shared with WDs \cite{YS2022TPDS}.
For MEC, there are two computation offloading policies, i.e., the binary offloading policy and the partial offloading policy~\cite{QXY2021TPDS}, \cite{WZL2023JSAC}. The former is appropriate for indivisible computing tasks, where each task is either computed locally at WDs (i.e., local computing mode) or entirely offloaded to the MEC server for computing (i.e., edge computing mode). The latter is appropriate for arbitrarily divisible computing tasks, where each task is divided into two parts. One part is computed locally at WDs, and the other part is offloaded to the MEC server for computing. 
{Works \cite{FG2020ICC} and \cite{AP2023TMC} focused on the partial offloading policy in a multi-user multi-server MEC environment, formulated a non-cooperative game, and proved the existence of the Nash equilibrium.}

Energy supply is a key factor impacting the computing performance, such as the computing delay and computing rate, and the offloading decisions, such as computing mode selections under the binary offloading policy and the offloading volume under the partial offloading policy. However, most WDs in IoT are powered by batteries with finite capacity. Frequent battery replacement is extremely cost or even impractical in hard-to-reach locations, which limits the lifetime of WDs. To break this limitation, wireless power transfer (WPT) technology \cite{DHP2022TMC,WY2022TCOM}, which realizes wireless charging of WDs by using hybrid access points (HAPs) or energy access points (EAPs) to broadcast radio frequency (RF) signals, is widely believed as a viable solution due to its advantages of stability and controllability in energy supply~\cite{LXY2023TWC,Lhh2024TVT}. As such, wireless powered mobile edge computing (WP-MEC) has recently aroused increasing attention~\cite{Deng2022TMC} since it combines the advantages of MEC and WPT technologies, i.e., enhancing computing capabilities of WDs while providing sustainable energy supply.

\begin{table*}[h]
    \caption{Comparison between this paper and the related works.}
	\centering
	\small
	\renewcommand\arraystretch{1.2}
	\begin{tabular}{|p{0.77cm}<{\centering}p{2.12cm}<{\centering} p{0.98cm}<{\centering} p{1.21cm}<{\centering} p{1.36cm}<{\centering} p{1.925cm}<{\centering}p{2.7cm}<{\centering}p{3.2cm}<{\centering}|}
      \hline
      \small\textbf{Ref.} & \small\textbf{Goal} & \small\textbf{Energy supply} & \small\textbf{Multiple servers} & \small\textbf{Binary offloading} & \small\textbf{Long-term optimization}  & \small\textbf{Computation data demand of network} & \small\textbf{Distributed computation offloading}\\
      \hline
      \hline
      \cite{BSZ2018TWC}  & \footnotesize computation rate   & $\surd$ & & $\surd$ & & & $\surd$  \\ \hline
	  \cite{ZFH2020TWC}  & \footnotesize energy efficiency  & $\surd$ & & $\surd$ & & &     \\ \hline
      \cite{CX2023TCOM}  & \footnotesize energy provision   & $\surd$ & &$\surd$ &$\surd$ & &   \\ \hline
      \cite{WF2020TWC}   & \footnotesize energy provision   & $\surd$ & & &$\surd$ & &   \\ \hline
      \cite{ZS2023IoT}   & \footnotesize energy provision   & $\surd$ & & & & &     \\ \hline
      \cite{YYH2022TVT}  & \footnotesize energy efficiency  & $\surd$ & & & & &     \\ \hline
      \cite{DJ2024TMC}   & \footnotesize energy efficiency  & $\surd$ & & & & $\surd$  &   \\ \hline
	  \cite{BSL2023TCOM}  & \footnotesize computation rate  & $\surd$ & $\surd$ & $\surd$ & & &     \\ \hline
      \cite{WXJ2023DCO}  & \footnotesize computation delay  & $\surd$ & $\surd$ & $\surd$ &$\surd$& &$\surd$ \\ \hline
      \cite{WF2018TWC}   & \footnotesize energy provision   & $\surd$ & & & & &   \\ \hline
      \footnotesize Ours & \footnotesize energy provision   & $\surd$ & $\surd$ &  $\surd$ & $\surd$ & $\surd$ & $\surd$ \\ \hline
	  \multicolumn{4}{l}{\small $\surd$ denotes the existence of the feature.}\\
	\end{tabular}
	\label{COMPARISON}
\end{table*}

Due to the coupling of the energy supply and the communication/computation demands of WDs, the critical issue in WP-MEC networks is how to reasonably allocate energy resource and make offloading decisions, so as to optimize various network performance. Consequently, many works~\cite{BSZ2018TWC,ZFH2020TWC,CX2023TCOM,PJ2021TVT,WF2020TWC,ZS2023IoT,YYH2022TVT,DJ2024TMC} have been done to address the issue.
Under the binary offloading policy,
Bi \emph{et al.} \cite{BSZ2018TWC} maximized the weighted sum computation rate of WDs in a multi-user WP-MEC network by jointly optimizing WDs' computing mode selections and the time allocation between WPT and data offloading.
Under the max-min fairness criterion, Zhou \emph{et al.} \cite{ZFH2020TWC} maximized the computation efficiency of a multi-user WP-MEC network by jointly optimizing the duration of the WPT phase, the CPU frequency, the time allocation and offloading power of WDs.
While Chen \emph{et al.} \cite{CX2023TCOM} considered a WP-MEC network where a multiple-antenna BS serves multiple WDs, and proposed an augmented two-stage deep Q-network (DQN) algorithm to minimize the average energy requirement of the network.
Under the partial offloading policy,
Park \emph{et al.} \cite{PJ2021TVT} investigated a WP-MEC with the simultaneous wireless information and power transmission (SWIPT) technique, and minimized the computation delay by jointly optimizing the data offloading ratio, the data offloading power, the CPU frequency of the WD and the power splitting ratio.
In a single-user WP-MEC network consisting of a multi-antenna EAP, a MEC server and a WD, Wang \emph{et al.} \cite{WF2020TWC} minimized the transmission energy consumption of the EAP during the WPT phase by jointly optimizing the energy allocation during the WPT phase at the EAP and the data allocation for offloading at the WD.  
{In a two-user WP-MEC network with the nonorthogonal multiple access (NOMA) protocol, Zeng \emph{et al.} \cite{ZS2023IoT} also minimized the transmission energy consumption of the EAP, similar to \cite{WF2020TWC}, under the energy and computing delay constraints, and proposed an iterative algorithm to solve it.
Different from the above works, works \cite{YYH2022TVT} and \cite{DJ2024TMC} focused on the backscatter-assisted WP-MEC network, where WDs harvest energy for local computing and data offloading through active transmission and backscatter communication. Considering the limited computation capacity of the MEC server, Ye \emph{et al.} \cite{YYH2022TVT} respectively maximized the computation energy efficiency and the total computation bits by proposing two resource allocation schemes. By leveraging the NOMA protocol to enhance backscatter communication, Du \emph{et al.} \cite{DJ2024TMC} maximized the computation energy efficiency.} 
The aforementioned works \cite{BSZ2018TWC,ZFH2020TWC,CX2023TCOM,PJ2021TVT,WF2020TWC,ZS2023IoT,YYH2022TVT,DJ2024TMC}, however, only focused on the WP-MEC networks with a single HAP, which makes it difficult to efficiently  process the massive amount of the computation data offloaded from a large number of WDs.

Recently, few works~\cite{BSL2023TCOM}, \cite{WXJ2023DCO} have studied the WP-MEC networks with multiple HAPs, which are practical for large-scale IoT.
Specifically, with the goal of computation rate maximization for a single time slot, Zhang \emph{et al.} \cite{BSL2023TCOM} first obtained the near-optimal offloading decisions by proposing an online deep reinforcement learning (DRL)-based algorithm, and then optimized the time allocation by designing a Lagrangian duality-based algorithm.
While Wang \emph{et al.} \cite{WXJ2023DCO} focused on the long-term average task completion delay minimization problem, and proposed an online learning algorithm implemented distributively for HAPs to learn both the duration of the WPT phase and offloading decisions at each time slot.
Actually, besides the computation rate \cite{BSZ2018TWC}, \cite{BSL2023TCOM}, computation efficiency \cite{ZFH2020TWC}, \cite{YYH2022TVT}, and computation delay \cite{PJ2021TVT}, \cite{WXJ2023DCO}, the energy provision is also a very important metric for evaluating the design of the WP-MEC networks~\cite{WF2018TWC}. However, as far as we know, the energy provision minimization of the WP-MEC networks with multiple HAPs has seldom been studied. Although works~\cite{CX2023TCOM} and \cite{WF2020TWC} have studied it in the WP-MEC networks with one HAP, 
the design in \cite{CX2023TCOM} and \cite{WF2020TWC} can not be applied in the WP-MEC networks with multiple HAPs due to the complex association
between WDs and HAPs.
 
{To fill this gap, we study the long-term energy provision minimization problem of a multi-HAP WP-MEC network in a dynamic environment, where WDs harvest energy from the RF signals of HAPs, and then compute or offload the computation data under the binary offloading policy.}  Besides the energy constraint, the computing delay and computation data demand constraints should be satisfied in order to ensure the computing performance of the network. The computing delay constraint ensures that the generated computation data at each time slot is computed either locally or remotely at HAPs within the acceptable duration. 
The computation data demand constraint ensures that the total amount of the processed computation data at each time slot is no smaller than the required computation data demand, which is in accordance with the real demand.
Since the amount of computation data generated by WDs and the channel gains between HAPs and WDs are uncertain in a dynamic network environment, optimization methods such as convex optimization and approximation method are difficult to well address the problem. Fortunately, DRL has been demonstrated in the literature as a more flexible and robust approach to adapt the MEC offloading decisions and the resource allocation by interacting with the dynamic network environment \cite{ZH2022HMAD}, \cite{2024tvtjoint}. Hence we exploit the DRL approach to address the problem. A straightforward implementation of the DRL approach is employing a centralized agent at HAPs to collect all network information and then adapt the actions for HAPs and WDs. However, with the increasing number of HAPs/WDs, the state space and action space increase explosively, resulting in long training time and poor performance. To address this dilemma, adopting a distributed computation offloading framework is a promising solution. {We summarize the differences between this paper and the related works in TABLE \ref{COMPARISON}, so as to highlight the novelty of this paper.}
The main contributions are summarized as follows.
\begin{itemize}
	\item
	In order to pursue a green computing design for a multi-HAP WP-MEC network, we formulate an energy provision minimization problem by jointly optimizing the transmit power of HAPs, the duration of the WPT phase,  the offloading decisions of WDs, the time allocation for offloading and the CPU frequency for local computing, subject to the energy, computing delay and computation data demand constraints. The formulated non-convex mixed integer programming (MIP) problem is very challenging to be tackled by proving it for a single time slot is NP-hard.
	\item
	To efficiently tackle the non-convex MIP problem in a distributed manner,  we decompose it into three subproblems, and then propose a two-stage multi-agent DRL-based distributed computation offloading (TMADO) framework to solve them.  The main idea is that the high-level agent residing in all HAPs is responsible for solving the first subproblem, i.e., optimizing the transmit power of HAPs and the duration of the WPT phase, while the low-level agents residing in WDs are responsible for solving the second and third subproblems, i.e., the offloading decisions of WDs, the time allocation for offloading and the CPU frequency for local computing. In a distributed way, each WD optimizes its offloading decision, time allocation for offloading, and CPU frequency.
	\item Simulation results validate the superiority of the proposed TMADO framework in terms of the energy provision minimization compared with comparison schemes.
	It is observed that when the number of HAPs/WDs reaches a certain value, the scheme with only edge computing mode is better than that with only local computing mode in terms of the energy provision minimization due to the reduced average distance between HAPs and WDs. It is also observed that, with the purpose of minimizing energy provision of HAPs, the WDs with high channel gains are prone to select local computing mode, and vice versa.
\end{itemize}

The rest of this paper is organized as follows.
In Section II, we introduce the system model of the WP-MEC network with multiple HAPs.
In Section III, we formulate the energy provision minimization problem.
In Section IV, we present the proposed TMADO framework.
In Section V, we present the simulation results.
Section VI concludes this paper.


\section{System Model}\label{systemmodel}

\subsection{Network Model}\label{networkmodel}
  
\begin{figure}[t]
	\centering
	\includegraphics[width=0.48\textwidth]{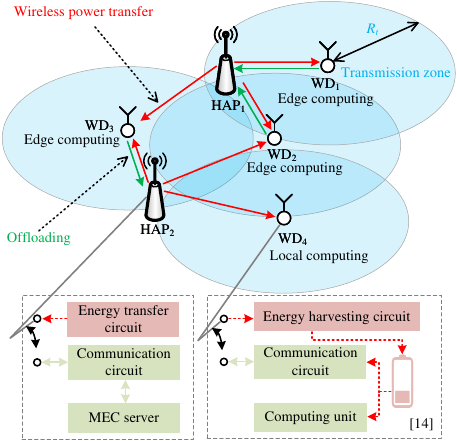}
	\caption{An example of the multi-HAP WP-MEC network.}
	\label{Fig.system_model1}
\end{figure}
     
\begin{figure}[t]
	\centering
	\includegraphics[width=0.48\textwidth]{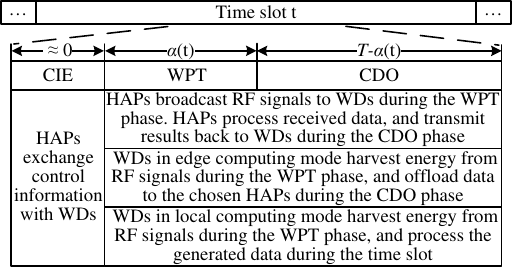}
	\caption{The time slot structure.}
	\label{Fig.system_model2}
\end{figure}

      
As shown in Fig. \ref{Fig.system_model1}, we study a WP-MEC network, where $M$ one-antenna HAPs, and $N$ one-antenna WDs coexist. Let  $\mathcal{M} = \{1,2,...,M \}$  denote the set of HAPs, and $\mathcal{N} = \{1,2,...,N \}$ denote the set of WDs. Equipped with one MEC server, each HAP broadcasts the RF signal to WDs through the downlink channel, and receives the computation data from WDs through the uplink channel. Equipped with one battery with capacity $E_b$, each WD harvests energy from RF signals for local computing or offloading the computation data to the chosen HAP. We consider that WDs adopt the harvest-then-offload protocol \cite{ZFH2020TWC}, i.e., WDs harvest energy before offloading data to the HAPs.
In accordance with the real demand, we define the transmission zone as a circle centered at the WD with radius $R_{t}$. The WD could offload its computation data to the HAPs within the corresponding transmission zone.

To process the generated computation data in the WP-MEC network, we consider that WDs follow the binary offloading policy, i.e., there are two computing modes for WDs. The WD in local computing mode processes the computation data locally by utilizing the computing units, and the WD in edge computing mode offloads the computation data to the chosen HAP through the uplink channel. After receiving the computation data, the HAP processes the received computation data, and transmits computation results back to the WD through the downlink channel.

As illustrated in Fig. \ref{Fig.system_model2}, the system time is divided into $\mathcal{T}$ time slots with equal duration $T$. {Time slot $t$ is divided into the control information exchanges (CIE) phase with negligible duration due to small-sized control information \cite{SQ2014TIE}, the WPT phase with duration $\alpha(t)$, and the computation data offloading (CDO) phase with duration $T-\alpha(t)$. 
During the CIE phase, the control information exchanges between HAPs and WDs complete.
During the WPT phase, HAPs broadcast RF signals to WDs through the downlink channels, and WDs harvest energy from RF signals of HAPs simultaneously.
During the CDO phase, the WDs in edge computing mode offload the computation data to the chosen HAPs through the uplink channels. While the WDs in local computing mode process the computation data during the time slot, since the energy harvesting circuit and the computing unit could work simultaneously \cite{ZFH2020TWC,DJ2024TMC}, as shown in the WD's circuit structure of Fig. \ref{Fig.system_model1}.}  At the beginning of time slot $t$, $N$ WDs generate computation data $\mathcal{D}(t) = [D_1(t), ..., D_N(t) ]$, where $D_1(t), ..., D_N(t)$ respectively denote the amount of the computation data generated by WD$_1$,  $...$, WD$_N$ at time slot $t$, and WDs transmit the state information about the amount of the generated computation data and the amount of the energy in the batteries to the HAPs. Then the HAPs follow the proposed TMADO framework to output the feasible solutions of the offloading decisions, broadcast them to WDs, and conduct WPT with duration $\alpha(t)$. According to the received feasible solutions of the offloading decisions, each WD independently makes the optimal offloading decision, i.e., it processes the generated computation data locally during the current time slot, or offloads the computation data to the chosen HAP during the CDO phase. 
{To avoid potential interference, the time division multiple access (TDMA) protocol is employed by WDs that offload the computation data to the same HAP at the same time slot, while the frequency division multiple access (FDMA) protocol is employed by WDs that offload the computation data  to different HAPs. As the MEC server of the
HAP has high CPU frequency \cite{WF2018TWC}, there is no competition among WDs for the edge resource.}
Similarly, as the MEC server of the HAP has high CPU frequency \cite{WF2018TWC} and the computation results are small-sized  \cite{ZFH2020TWC}, we neglect the time spent by HAPs on processing data and transmitting the computation results back to the WDs.

\subsection{Channel Model}
For the channels in the WP-MEC network, we adopt the free-space path loss model for the large-scale fading, and adopt the Rayleigh fading model \cite{2024TMCon} for the small-scale fading.  
 Let ${h}_{n,m}(t)$  denote the channel gain between WD$_n$ and HAP$_m$ at time slot $t$ as
\begin{equation}
	{h}_{n,m}(t)={\sigma }_{L,n,m} {\left|{\sigma}_{S,n,m}(t)\right|}^{2},
\end{equation}
where ${\sigma}_{L,n,m}$ denotes the large-scale fading component, and ${\sigma}_{S,n,m}(t)$ denotes the small-scale fading component at time slot $t$. The uplink channel and downlink channel are considered to be symmetric \cite{BSZ2018TWC}, i.e., the channel gain of the uplink channel equals that of the corresponding downlink channel.
We consider that channel gains remain unchanged within each time slot, and vary from one time slot to another. 


\subsection{Energy Harvesting Model}

During the WPT phase, WDs harvest energy from the RF signals broadcasted by HAPs. The amount of the energy harvested by WDs depends on the transmit power of the HAPs, the distance between WDs and HAPs, and the duration of the WPT phase. 
 We adopt the linear energy harvesting (EH) model, and formulate the amount of the energy harvested by WD$_n$ at time slot $t$ as
 
\begin{equation}\label{E_{h,n}}
	E_{h,n}(t) = \sum _{m=1}^{M} \mu {P}_{{h},m}(t){h}_{n,m}(t){\alpha(t)},
\end{equation}
where $  \mu \in (0,1)  $ denotes the EH efficiency,  ${P}_{{h},m}(t) \in [0, {P}_{\max}] $ denotes the transmit power of HAP$_m$ at time slot $t$, and ${P}_{\max}$ denotes the maximum transmit power of HAPs.

The amount of the available energy in WD$_n$ at time slot $t$, i.e., the sum of the amount of the initial energy in WD$_n$ and the amount of the energy harvested by WD$_n$ at time slot $t$, is formulated as
\begin{equation}\label{Eini}
	E_{n}(t) =\text{min}\{E_{i,n}(t) + E_{h,n}(t), E_{b} \},
\end{equation}
where $E_{i,n}(t)$ denotes the amount of the initial energy in WD$_n$ at time slot $t$. At the beginning of the first time slot, the initial energy in WDs equals zero, i.e., $E_{i,n}(0) = 0, \forall n \in  \mathcal{N}$.
As shown in (\ref{Eini}), $\text{min}\{E_{i,n}(t) + E_{h,n}(t), E_{b} \}$ ensures that the amount of the available energy in WD$_n$ does not exceed the battery capacity.
Then the amount of initial energy in WD$_n$ at time slot $t+1$ updates as
 \begin{equation}\label{EiniUpdate}
	E_{i,n}(t+1) =\text{min}\{E_{i,n}(t) + E_{h,n}(t), E_{b} \} - E_{{w},n}(t),
\end{equation}
where $E_{{w},n}(t)$ denotes the amount of the energy consumed by WD$_n$ at time slot $t$.

\subsection{Offloading Decisions}
According to the computing modes, WDs are divided into nonoverlapping sets $\mathcal{N}_m(t) $, $\mathcal{N}_0(t) $, and $\mathcal{N}_{\text{-1}}(t)  $.
Let $\mathcal{N}_m(t), m \in \mathcal{M}$ denote the set of WDs that process the computation data by offloading it to HAP$_m$ at time slot $t$, $\mathcal{N}_0(t) $ denote the set of WDs that process the computation data locally at time slot $t$, and $\mathcal{N}_{\text{-1}}(t) $ denote the set of WDs that fail to process the computation data at time slot $t$.
 If the available energy in WD is not enough for local computing in (\ref{E_{l,n}}) or edge computing in (\ref{E_{o,n}}), WD fails to process the computation data at time slot $t$. 
The WDs failing to process the computation data at time slot $t$ are not in the feasible solutions of the offloading decisions $(\mathcal{N}- \mathcal{N}_{\text{-1}}(t))$ from HAPs.
{Considering the quality of service requirement for the WDs in $\mathcal{N}_{\text{-1}}(t)$, the WDs in $\mathcal{N}_{\text{-1}}(t)$ directly drop the computation data without retransmitting, and accumulate the energy in the batteries so as to successfully process the computation data of the following time slots. $\mathcal{N}_{\text{-1}}(t) $ varies across time slots since WDs with sufficient harvested energy process the generated computation data locally, or offload it to the chosen HAPs.
}
Then $\mathcal{N}_m(t) $, $\mathcal{N}_0(t)$, and $\mathcal{N}_{\text{-1}}(t) $ satisfy
\begin{equation}
 	\mathcal{N}  = \bigcup ^{M}_{m= -1}{\mathcal{N}_m (t)} ,
\end{equation}
where $ \mathcal{N}_x(t)  \cap \mathcal{N}_y(t)  = \phi $, $x, y \in \{-1,...,M\}$, $x\neq y $, and $\phi $ denotes the empty set.

\subsection{Local Computing Mode}
According to the received feasible solutions of the offloading decisions from HAPs, when WD$_n$ chooses local computing mode at time slot $t$, i.e., $n \in  \mathcal{N}_{0}(t)$, it processes the generated computation data of $ D_{n}(t) $ bits locally during the whole time slot. Let $f_n(t)$ with $f_{n}(t) \in (0, f_{\max}] $ denote the CPU frequency of WD$_n$ by using dynamic voltage and frequency scaling technique \cite{BS2014SP} at time slot $t$, where $f_{\max}$ denotes the maximum CPU frequency of WDs. Let $C_n > 0$ denote the number of CPU cycles required by WD$_n$ to process 1-bit data. The computing delay constraint, i.e., the delay of WD$_n$ to process the generated computation data $D_{n}(t)$ at time slot $t$, denoted by $\tau_{l,n}(t)$, does not exceed the duration of the time slot $T$, is formulated as

\begin{equation}\label{tloc1}
	\tau_{l,n}(t) = \frac{{C}_{n}{D}_{n}(t)}{f_{n}(t)} \le T , \forall n \in \mathcal{N}_{0}(t) .
\end{equation}
Since energy is the product of power and time, the amount of the energy consumed by WD$_n$ for local computing at time slot $t$ is the product of the CPU power of WD$_n$, denoted by $P_{\text{cpu},n}(t)$, and the delay of WD$_n$ to process the generated computation data, i.e., $\tau_{l,n}(t)$. According to \cite{CLS1994COMPCON} and \cite{GST2019TMC}, the CPU power of WD$_n$ at time slot $t$ is expressed as 
\begin{equation} \label{ElocO1}
 P_{\text{cpu},n}(t) =C_{L,n} V_n^2(t) f_n(t),
\end{equation}
where $C_{L,n}$ denotes the switched load capacitance of WD$_n$, and $V_n(t)$ denotes the CPU voltage of WD$_n$ at time slot $t$.
As pointed out by \cite{GST2019TMC}, when the CPU operates at the low voltage limit, which is in accordance with the real-world WDs, the CPU frequency is approximately linear with the CPU voltage. Consequently, $P_{\text{cpu},n}(t)$ in (\ref{ElocO1}) can be reformulated as $k_n f_n^3(t)$, where $k_n$ satisfying $C_{L,n} V_n^2(t) = k_n f_n^2(t)$ is called the effective switched load capacitance of WD$_n$ \cite{LTH2023ToN}. Then the amount of the energy consumed by WD$_n$ for local computing at time slot $t$ is expressed as 
\begin{equation} \label{Eloc1}
E_{l,n}(t)\!={P_{\text{cpu},n}(t) \tau_{l,n}(t)} ={k_n f^3_{n}(t) \tau_{l,n}(t)},  \forall n \in  \mathcal{N}_{0}(t),
\end{equation}
which has been extensively used in the works related to MEC networks \cite{BSZ2018TWC, ZFH2020TWC, YYH2022TVT, DJ2024TMC, LTH2023ToN}.



{Since WDs in local computing mode are powered by the energy harvested during the WPT phase and that stored in the batteries, the energy constraint for WD$_n$ in local computing mode, i.e., the amount of the energy consumed by WD$_n$ for local computing at time slot $t$ is no larger than the available energy in WD$_n$ at time slot $t$, is formulated as \cite{WF2018TWC}}
\begin{equation}\label{E_{l,n}}
E_{l,n}(t)  \le E_{n}(t)  ,  \forall n \in  \mathcal{N}_{0}(t).
\end{equation}

\subsection{Edge Computing Mode}

According to the received feasible solutions of the offloading decisions from HAPs, when WD$_n$ chooses edge computing mode at time slot $t$, it offloads the computation data to HAP$_m$, i.e., $n \in \mathcal{N}_{m}(t)$, $m \in \mathcal{M}$. Let $\tau_{o,n,m}(t)$ denote the duration that WD$_n$ offloads the computation data of $ D_{n}(t) $ bits to HAP$_m$ at time slot $t$. 
Then $\tau_{o,n,m}(t)$ and $\alpha(t)$ satisfy
\begin{equation}\label{tau_{o,n,m}}
	\sum _{n \in {\mathcal{N}_{m}(t)}}\tau_{o,n,m}(t)+ \alpha(t) \le T , \forall  m \in \mathcal{M}.
\end{equation}
Based on Shannon's formula \cite{LH2023TMC}, to ensure that WD$_n$ in edge computing mode successfully offloads the computation data to HAP$_m$ during the CDO phase, $\tau_{o,n,m}(t)$ satisfies
\begin{equation}\label{tau_{n,m}}
	\tau_{o,n,m}(t) \ge \frac{v {D}_{n}(t) }{B {\mathrm{log}}_{2}\left(1+\frac{P_{n} {h}_{n,m}(t)}{{N}_{0}}\right)}
	, \forall n \in \mathcal{N}_{m}(t), m \in \mathcal{M} ,
\end{equation}
where $B$ denotes the uplink bandwidth of each HAP, and ${v} \geq 1$ represents the communication overhead including the encryption and data header costs \cite{BSZ2018TWC}. $P_{n}$ denotes the transmit power of WD$_n$, and ${N}_{0}$ denotes the power of the additive white Gaussian noise (AWGN).
{Besides, there is no interference between multiple WDs in (\ref{tau_{n,m}}) due to the TDMA protocol.}
Let $E_{o,n}(t)$ denote the amount of the energy consumed by WD$_n$ for transmitting the computation data of $D_n$ bits to HAP$_m$ at time slot $t$ as
\begin{equation}\label{E_{o,n_0}}
	E_{o,n}(t)= (P_ {n}+ P_{c,n} )	\tau_{o,n,m}(t)   , \forall n \in \mathcal{N}_{m}(t),   m \in \mathcal{M},
\end{equation}
where $ P_{c,n}$ denotes the circuit power of WD$_n$. During the CDO phase, the energy constraint for WD$_n$ in edge computing mode, i.e., the amount of the energy consumed by WD$_n$ for edge computing is no larger than that of the available energy in WD$_n$ at time slot $t$, is formulated as
\begin{equation}{\label{E_{o,n}}}
	E_{o,n}(t) \le E_{n}(t)  , \forall n \in \mathcal{N}_{m}(t),   m \in \mathcal{M} .
\end{equation}
Then the amount of the energy consumed by WD$_n$ at time slot $t$ is defined as
\begin{equation}
	{E_{{w},n}(t)} =
	\begin{cases}
	E_{o,n}(t),&{\text{if}}\ n \in \mathcal{N}_{m}(t),   m \in \mathcal{M} ;\\
	E_{l,n}(t),&{\text{if}}\ n \in \mathcal{N}_{0}(t) ;\\
	 {0 ,}&{\text{otherwise.}}
	\end{cases}
\end{equation}
If the WD fails to process the computation data under the TMADO framework at time slot $t$, its computation data would not be scheduled \cite{WXJ2023DCO}, and $E_{{w},n}(t)$ equals $0$.

The amount of the energy consumed by HAP$_m$ at time slot $t$, denoted by $E_{{h},m}(t)$, is formulated as
\begin{equation}\label{E_HAP}
E_{{h},m}(t) = E_{1,m}(t) + E_{2,m}(t).
\end{equation}
{(\ref{E_HAP}) indicates that the energy provision by HAPs consists of the amount of the energy consumed by HAPs for broadcasting RF
signals and the amount of the energy consumed by HAPs for processing the received computation data from WDs.}
During the WPT phase of time slot $t$, the amount of the energy consumed by HAP$_m$ for broadcasting the RF signal, denoted by $E_{1,m}(t)$, is formulated as
\begin{equation}\label{E_{1,m}}
	E_{1,m}(t) = \alpha(t) {P}_{{h},m}(t), \forall  m \in \mathcal{M} .
\end{equation}
{During the CDO phase of time slot $t$, the amount of the energy consumed by HAP$_m$ for processing the received computation data of $ \sum_{n \in {\mathcal{N}_{m}(t)}} D_{n}(t)$ bits from the WDs in $\mathcal{N}_m(t)$, denoted by $E_{2,m}(t)$, is formulated as \cite{BT2021TWC} 
 \begin{equation}\label{E_{2,m}}
 		E_{2,m}(t) = \sum_{n \in {\mathcal{N}_{m}(t)}}{e_m D_{n}(t)} , \forall m \in \mathcal{M} ,
 \end{equation}
 where $e_m$ represents the energy consumption of HAP$_m$ for processing per offloaded bit.}

\section{Problem Formulation}

In this section, we formulate the long-term energy provision minimization problem in the multi-HAP WP-MEC network as $\mathbf{P0}$. Then we adopt Lemmas \ref{Lemma1-2} and \ref{Lemma1} to prove that $\mathbf{P0}$ for a single time slot is NP-hard, which makes $\mathbf{P0}$ much more perplexed and challenging.

The total amount of the energy provision by HAPs at time slot $t$, denoted by $\Psi(t)$, is expressed as
\begin{equation}\label{Psi1}
	\Psi(t) =  \sum_{m=1}^{M} 	E_{{h},m}(t) .
\end{equation}
We aim to minimize $\Psi(t)$ in the long term through optimizing the transmit power of HAPs $\boldsymbol{{P}_{{h}}(t)} = [{P}_{{h},1}(t), ..., {P}_{{h},M}(t)]$, the duration of the WPT phase $\alpha(t)$, the offloading decisions $\boldsymbol{\mathcal{X}(t)} =\{\mathcal{N}_{0} (t), ..., \mathcal{N}_M (t)\}$, the time allocated to WDs for offloading the computation data to HAPs $\boldsymbol{\tau_{o}(t)} = [\tau_{o,1,1}(t), ..., \tau_{o,1,M}(t), \tau_{o,2,1}(t), ..., \tau_{o,2,M}(t), ..., \tau_{o,N,M}(t)]$, and the CPU frequencies of WDs $\boldsymbol{f(t)} = [f_{1}(t), ..., f_{N}(t) ] $
as
\begin{subequations}
	\label{P0}
	\begin{flalign}
		\begin{split}
			\mathbf{P0}&: \underset{\boldsymbol{{P}_{{h}}(t)},\alpha(t),\boldsymbol{\mathcal{X}(t)}, \boldsymbol{\tau_{o}(t)},\boldsymbol{f(t)}}    {\text{min}}~
			\sum_{t=0}^{\mathcal{T}-1} \Psi(t) \label{Objective:AP} 
		\end{split}
		\\
		&\mathrm{s.t.}~ \sum_{n \in (  \mathcal{N}- \mathcal{N}_{\text{-1}}(t)   )}  D_{n}(t) \geq D_{th}, ~
		\label{Const:Ddemand}\\
		& ~~~~~ 0 \le  \frac{{C}_{n}{D}_{n}(t)}{f_{n}(t)} \le T,~~~\forall n \in \mathcal{N}_{0}(t),   \label{Const:LocT} \\
		&~~~~~0 \leq {\tau}_{o,n,m}(t) \leq T, ~~~\forall n \in \mathcal{N}_{m}(t),   m \in \mathcal{M}, \label{Const:OffT_tau}\\
		&~~~~~0 \leq \alpha(t) \leq T, \label{Const:OffT_alpha} \\
		&~~~~\sum _{n \in {\mathcal{N}_{m}(t)}}{{\tau}}_{o,n,m}(t)+\alpha(t)\le T,~   ~~~\forall m\in \mathcal{M}, \label{Const:OffT} \\
		&~~~~~ 0 \le {P}_{{h},m}(t) \le {P}_{\max},~~~ \forall m \in \mathcal{M},  \label{Const:PHAP} \\
		&~~~~~k_n f^3_{n}(t) \tau_{l,n}(t) \leq E_{n}(t),~~~\forall n \in \mathcal{N}_{0}(t),   \label{Const:LocE} \\
		& ~~~~~\left(P_{n}+P_{c,n}\right) \tau_{o, n,m}(t) \leq E_{n}(t), \nonumber\\ &~~~~~~~~~~~~~~~~~~~~~~~~~~~~\forall n \in \mathcal{N}_{m}(t),   m \in \mathcal{M},\label{Const:OffE}  \\
		& ~~~~~ 0 < f_{n}(t) \leq f_{\max},~~~ \forall n \in \mathcal{N}_{0}(t),\label{Const:Locf} \\
		& ~~~~~ E_{n}(t) =\text{min}\{E_{i,n}(t) \! + \!E_{h,n}(t), E_{b} \},\forall n \in \mathcal{N}. \label{Const:Battery}
	\end{flalign} 
\end{subequations} 
In $\mathbf{P0}$, (\ref{Const:Ddemand}) represents the computation data demand constraint that the total amount of the processed computation data at each time slot is no smaller than the computation data demand $D_{th}$. (\ref{Const:LocT}) represents the computing delay constraint for each WD in local computing mode. (\ref{Const:OffT_tau}), (\ref{Const:OffT_alpha}), and (\ref{Const:OffT}) respectively represent that the duration of offloading the computation data to HAP$_m$, the duration of the WPT phase, and the sum of the total duration of the offloaded computation data to HAP$_m$ and the duration of the WPT phase do not exceed the duration of the time slot. (\ref{Const:PHAP}) ensures that the transmit power of each HAP does not exceed the maximum transmit power of HAPs. {(\ref{Const:LocE}) represents the energy constraint for each WD in local computing mode \cite{WXJ2023DCO}.} (\ref{Const:OffE}) represents that the amount of the energy consumed by the WD in edge computing mode is no larger than that of the available energy in the WD.  (\ref{Const:Locf}) ensures that the CPU frequency of each WD does not exceed the maximum CPU frequency. {(\ref{Const:Battery}) ensures that the amount of the available energy in each WD, which is equivalent to the sum of the amount of the initial energy in each WD and that of the energy harvested by each WD, does not exceed the battery capacity.}


In the following, we provide a detailed explanation of the NP-hardness of the formulated non-convex MIP problem.
As defined in \cite{BS2012SORMS}, we provide the definition of the non-convex MIP problem as 
\begin{subequations}
	\label{MIP-R}
	\begin{flalign}
		\begin{split}
			& \underset{x,y}    {\text{min}}~
			f_0(x, y)
		\end{split}
		\\
		&\mathrm{s.t.}~  f_i(x, y) \leq 0 , ~i=\{1, \ldots, m\},
		\label{Const:MIP}\\
		& ~~~~~ x \in \mathbb{Z}_{+}^{n_1},  y \in \mathbb{R}_{+}^{n_2},   \label{Const:MIPvaribles} 
	\end{flalign}
\end{subequations}
where $x$ denotes the vector of integer variables, $y$ denotes the vector of continuous variables, $f_0(x, y), f_1(x, y), ..., f_m(x, y)$ represent the arbitrary functions mapping from $\mathbb{Z}_{+}^{n_1} \times \mathbb{R}_{+}^{n_2}$ to the real numbers, $n_1>0$ denotes the number of integer variables, $n_2 \ge 0$ denotes the number of continuous variables, and $m \ge 0$ denotes the number of constraints. The MIP problem is considered as a general class of problems, consisting of the convex MIP problem and the non-convex MIP problem. The MIP problem is convex if $f_0(x, y), f_1(x, y), ..., f_m(x, y)$ are convex, and vice versa \cite{BS2012SORMS}. 
 
Then we explain the NP-hardness of the non-convex MIP problem in this paper. According to (\ref{Objective:AP})-(\ref{Const:Battery}), the formulated energy provision minimization problem for time slot $t$ that optimizes $\boldsymbol{{P}_{{h}}(t)}$, $\alpha(t)$,  $\boldsymbol{\mathcal{X}(t)}$, $\boldsymbol{\tau_{o}(t)}$, and  $\boldsymbol{f(t)}$, is
\begin{subequations}
	\label{P-1}
	\begin{flalign}
		\begin{split}
			& \underset{\boldsymbol{{P}_{{h}}(t)},\alpha(t),\boldsymbol{\mathcal{X}(t)}, \boldsymbol{\tau_{o}(t)},\boldsymbol{f(t)}}    {\text{min}}~ \!\!\!\!\sum_{m=1}^{M}(\alpha(t) {P}_{{h},m}(t) \!+ \!\!\!\!\!\! \sum_{n \in {\mathcal{N}_{m}(t)}}\!\!\!\!\!\!{e_m D_{n}(t)}) \label{objective:ap}
		\end{split}
           \\
		&\mathrm{s.t.}~ (\ref{Const:Ddemand})-(\ref{Const:Battery}).  \nonumber
	\end{flalign}
\end{subequations}
It is observed that (\ref{objective:ap}) includes integer variables $\boldsymbol{\mathcal{X}(t)}$ and the non-convex function resulting from the coupling relationship between  $\boldsymbol{{P}_{{h}}(t)}$ and $\alpha(t)$. According to the definition of the non-convex MIP problem, the formulated energy provision minimization problem for time slot $t$ is a non-convex MIP problem. Then we adopt Lemmas \ref{Lemma1-2}-\ref{Lemma1} to prove this non-convex MIP problem is NP-hard.

\begin{lemma}{\label{Lemma1-2}}
	Given {\rm $\boldsymbol{{P}_{{h}}(t)}$, $\alpha(t)$, $\mathcal{N}_\text{-1}(t)$, $\tau_{l,n}(t)$, $f_{n}(t)$} in the optimal solution of {\rm $\mathbf{P0}$} for time slot $t$, i.e., $\mathcal{T} = 1$, WD{\rm $_{n}$} that satisfies the constraints in (\ref{Const:Ddemand}), (\ref{Const:LocE}), and (\ref{Const:Locf}) at time slot {\rm $t$} chooses local computing mode, i.e., {\rm $ n \in \mathcal{N}_0(t) $}.
	\end{lemma}
	\begin{IEEEproof}
	Satisfying the constraints in (\ref{Const:Ddemand}), (\ref{Const:LocE}), and (\ref{Const:Locf}) indicates that WD{\rm $_{n}$} satisfies the computation data demand constraint, energy constraint, and CPU frequency constraint. Then $ n \in \mathcal{N}_0(t) $ is a feasible solution of $\mathbf{P0}$.
	{With given {\rm $\boldsymbol{{P}_{{h}}(t)}$, $\alpha(t)$, $\mathcal{N}_\text{-1}(t)$, $\tau_{l,n}(t)$, $f_{n}(t)$}, the value of $E_{1,m}$ in (\ref{E_{1,m}}) is determined. Based on (\ref{E_HAP})-(\ref{Psi1}), the energy provision by HAPs in $\mathbf{P0}$ with WD{\rm $_{n}$} in local computing mode is smaller than that with WD{\rm $_{n}$} in edge computing mode. This completes the proof.}
	\end{IEEEproof}

\begin{lemma}{\label{Lemma1}}
With given {\rm $\boldsymbol{{P}_{{h}}(t)}$, $\alpha(t)$, $\mathcal{N}_\text{-1}(t)$, $\boldsymbol{\tau_{o}(t)}$, and $\boldsymbol{f(t)}$, $\mathbf{P0}$ for time slot $t$} is NP-hard.
\end{lemma}

\begin{IEEEproof}
With given $\boldsymbol{{P}_{{h}}(t)}$, $\alpha(t)$, $\mathcal{N}_\text{-1}(t)$, $\boldsymbol{\tau_{o}(t)}$, and $\boldsymbol{f(t)} $, the value of $E_{1,m}(t)$ in (\ref{E_{1,m}}) is determined. Based on (\ref{E_HAP}), $\mathbf{P0}$ for time slot $t$ aims to minimize $\sum_{m=1}^{M} E_{2,m}(t)$ in (\ref{E_{2,m}}) and (\ref{Psi1}) by optimizing the offloading decisions $\boldsymbol{\mathcal{X}(t)} = \{\mathcal{N}_{0} (t), ..., \mathcal{N}_M (t)\}$. Based on (\ref{E_{2,m}}), we infer that $\sum_{m=1}^{M} E_{2,m}(t)$ is independent of $\mathcal{N}_0(t)$. Based on Lemma \ref{Lemma1-2},  $\mathcal{N}_0(t)$ in the optimal offloading solution $\boldsymbol{\mathcal{X}(t)}^*$ of $\mathbf{P0}$ for time slot $t$ is determined.

According to the aforementioned analysis, we transform $\mathcal{N}_m(t) \subset \boldsymbol{\mathcal{X}(t)}^*$ for time slot $t$ into the multiple knapsack problem as follows. Treat $M$ HAPs as $M$ knapsacks with the same load capacity $T-\alpha(t)$. Treat  $N$ WDs as $N$ items with the weight of item$_n$ equaling the duration that WD$_n$ offloads the computation data of $D_n(t)$ bits to the chosen HAP. The value of item$_n$ is equal to the amount of the energy consumed by the chosen HAP for processing the computation data of $ D_n(t)$ bits from WD$_n$. Then finding $\mathcal{N}_m(t) \subset \boldsymbol{\mathcal{X}(t)}^*$ is equivalent to finding the optimal item-assigning solution to minimize the total value of the items assigned to $M$ knapsacks. The multiple knapsack problem is NP-hard \cite{CACCHIANI2022105693}. This completes the proof.
\end{IEEEproof}
 
{As aforementioned, $\mathbf{P0}$ is a non-convex MIP problem, and we provide Lemmas \ref{Lemma1-2}-\ref{Lemma1} to demonstrate that this non-convex MIP problem is NP-hard in general.}
Therefore, it is quite challenging to solve $\mathbf{P0}$, especially in a distributed manner, i.e.,
each WD independently makes its optimal offloading decision according to the local observation. Each WD can not capture the state information of other WDs and the HAPs that are located outside the corresponding transmission zone.
\begin{figure}
	\centering
	\includegraphics[width=0.49\textwidth]{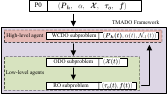}
	\caption{The flow chart of the proposed TMADO framework.}
	\label{Fig.flowchart}
\end{figure}

\section{TMADO Framework}

In essence, $\mathbf{P0}$ is a sequential decision-making question. As a power tool for learning effective decisions in dynamic environments, DRL is exploited to tackle $\mathbf{P0}$.
Specifically, as
shown in Fig. \ref{Fig.flowchart}, we propose the TMADO framework to decompose $\mathbf{P0}$ into three subproblems, i.e., WPT and computation data optimization (WCDO) subproblem, offloading decision optimization (ODO) subproblem, and resource optimization (RO) subproblem.
We specify the TMADO framework as follows.
\begin{itemize}

\item With state information of HAPs, that of WDs, and that of channel gains, the WCDO subproblem optimizes the transmit power of HAPs $\boldsymbol{{P}_{{h}}(t)}$, duration of the WPT phase $\alpha(t)$, and feasible solutions of the offloading decisions $(\mathcal{N}- \mathcal{N}_{\text{-1}}(t)) $ by HAPs, subject to the computation data demand constraint (\ref{Const:Ddemand}),  duration of the WPT phase constraint (\ref{Const:OffT_alpha}), and transmit power constraint (\ref{Const:PHAP}).


\item With the output of the WCDO subproblem  \{$\boldsymbol{{P}_{{h}}(t)}$, $\alpha(t)$, $\mathcal{N}_\text{-1}(t)$\},
and the local observations by each WD,
the ODO subproblem optimizes the offloading decision $\boldsymbol{\mathcal{X}(t)}$ by WDs, subject to the computing delay constraint (\ref{Const:OffT_tau}) and energy constraints (\ref{Const:LocE}), (\ref{Const:OffE}).
\item With the output of the WCDO subproblem \{$\boldsymbol{{P}_{{h}}(t)}$, $\alpha(t)$, $\mathcal{N}_\text{-1}(t)$\}, and the output of the ODO subproblem $\boldsymbol{\mathcal{X}(t)}$, the RO subproblem outputs the duration that WDs in edge computing mode offload the computation data to HAPs $\boldsymbol{\tau_{o}(t)}$, and the CPU frequencies $\boldsymbol{f(t)}$ of WDs in local computing mode
, subject to the computing delay constraints (\ref{Const:LocT}), (\ref{Const:OffT}), and CPU frequency constraint (\ref{Const:Locf}).
\end{itemize}
The HAPs are considered as a single high-level agent assisted by the server of cloud center in the WCDO subproblem. Each WD is considered as a low-level independent agent in the ODO subproblem.
Under the TMADO framework, the decision-making process is completed by both HAPs as the high-level agent and WDs as low-level agents due to the following reasons. 1) Completing the decision-making process by only the high-level agent faces challenges of large-scale output actions and poor scalability.
2) Completing the decision-making process by only low-level agents makes it difficult to determine the duration of the WPT phase and satisfy the computation data demand in (\ref{Const:Ddemand}). 
3) It is difficult to determine the actions output by only high-level agent and those output by only low-level agents due to the coupling relationship between the output of the WCDO subproblem and that of the ODO subproblem.

Based on aforementioned three reasons, we explore a hierarchical structure in the TMADO framework that consists of a single-agent actor-critic architecture based on deep deterministic policy gradient (DDPG), and a multi-agent actor-critic architecture based on independent proximal policy optimization (IPPO).
The HAPs first determine the high-level action through the DDPG algorithm. Then each low-level WD updates its own offloading decision through the IPPO algorithm. The states of the high-level agent are influenced by the low-level WDs' decisions, driving the HAPs to update their actions in the next decision epoch. 
{The TMADO framework performs the control information exchanges between HAPs and WDs during the CIE phase of each time slot as follows. 
1) HAPs and WDs transmit states to the high-level agent through the dedicated control channel by using the TDMA protocol. Specifically, HAPs transmit the states of the energy consumption to the high-level agent, and WDs transmit the states of data generation and initial energy to the high-level agent. 
2) The high-level agent combines the received states, obtains its action through the high-level learning model, and broadcasts its action to HAPs and WDs.
3) Each low-level agent receives its state, obtains its action through the low-level learning model, and then executes its action.}

The WCDO subproblem involves not only the integer variable (i.e., $\mathcal{N}_\text{-1}(t)$) but also the continuous variables (i.e., $\boldsymbol{{P}_{{h}}(t)}$ and $\alpha(t)$). The DDPG does well in searching for the optimal policy with continuous variables for a single agent \cite{DDPG}. Hence the DDPG is employed by the high-level agent at the HAPs to solve the WCDO subproblem.
Whereas the ODO subproblem involves only integer variables (i.e., $\boldsymbol{\mathcal{X}(t)}$), and the multi-WD offloading decisions need to be made by the multiple low-level agents at WDs. Different from other multi-agent DRL algorithms such as multi-agent deep deterministic policy gradient (MADDPG), multi-agent deep Q-learning network (MADQN) and multi-agent proximal policy optimization (MAPPO), IPPO estimates individual state value function of each agent without the interference from irrelevant state information of other agents, and then could achieve a better reward~ \cite{YYX2023ICDE}. Hence IPPO is employed by the low-level agents at WDs to solve the ODO subproblem.

\subsection{WCDO Subproblem}

We formulate the WCDO subproblem as a Markov decision process (MDP), represented by a tuple of the high-level agent $(\mathcal{S}^{h},  \mathcal{A}^{h}, \mathcal{R}^{h}, \gamma^{h} )$, where $\mathcal{S}^{h}$,  $\mathcal{A}^{h}$, $\mathcal{R}^{h}$, and $\gamma^{h}$ represent state, action, reward, and the discount factor, respectively. The high-level agent receives state $\boldsymbol{s}^{h}(t) \in \mathcal{S}^{h}$, selects action $\boldsymbol{a}^{h}(t) \in \mathcal{A}^{h}$ at time slot $t$, and receives reward $r^{h}(t) \in \mathcal{R}^{h}$ and state $\boldsymbol{s}^{h}(t+1)$.

\begin{itemize}

\item  State space: The global state of the WCDO subproblem at time slot $t$ is defined as
\begin{equation}{\label{sh}}
	\begin{split}
		\boldsymbol{s}^{h}(t) = \{ \boldsymbol{s}_p^{h}(t), \boldsymbol{s}_w^{h}(t), \boldsymbol{s}_c^{h}(t)\},
	\end{split}
\end{equation}
where $\!\boldsymbol{s}_p^{h}(t)\!\!\!\!=\!\!\!\!\{E^{\text{tot}}_{{h},1}(t),..., E^{\text{tot}}_{{h},M}(t) \}$ represents the state information of HAPs{\spaceskip=0.4em\relax, }
$\boldsymbol{s}_w^{h}(t)=\{D_{1}(t),..., D_{N}(t), E_{i,1}(t),..., E_{i,N}(t) \}$ represents that of WDs, and $\boldsymbol{s}_c^{h}(t)=\{{h}_{n,m}(t)| n\in \mathcal{N}, m\in \mathcal{M} \}$ represents the channel gains at time slot $t$.
$E^{\text{tot}}_{{h},m}(t)$ is the state information of HAP$_m$ that represents the total amount of the energy consumed by HAP$_m$ during time slots $[0,t-1)$.
$E_{i,n}(t)$ represents the amount of the initial energy in WD$_n$ at time slot $t$.

\item  Action space: At time slot $t$, the action space of HAPs is defined as
\begin{equation}{\label{A^h}}
	\begin{split}
		\!\mathcal{A}^{h}(t) \!\!  = \!\!\{ \alpha(t), {P}_{{h},1}(t),..., &{P}_{{h},M}(t),\\ & a_{{c},1}(t),...,a_{{c},N}(t)\} \!\!\in \!\mathcal{A}^{h},
	\end{split}
\end{equation}
where $a_{\text{c},n}(t)$ represents the energy provision cost of HAPs for the energy supply of WD$_n$ estimated by the high-level agent, i.e., the amount of the energy consumed by HAPs for WD$_n$ to harvest per joule energy.
To be specific, if the energy provision cost of HAPs for the energy supply of WD$_n$ is small, the probability that HAPs determine that WD$_n$ in local computing mode or edge computing mode processes the computation data is high. Otherwise, the probability that HAPs determine that WD$_n$ is not in the feasible solutions of the offloading decisions is high.
As $a_{\text{c},n}(t)$ is a continuous action, and the feasible solutions of the offloading decisions are discrete actions, we convert continuous action $a_{\text{c},n}(t)$ into discrete action, which represents the set of WDs to process the computation data, and obtain $\mathcal{N}_{\text{-1}}(t)$. To optimize the feasible solutions of the offloading decisions, the sub-action space $\mathcal{A}_c^{h}(t)$ is established in the ascending order of action $a_{\text{c},n}(t)$ as
\begin{equation}{\label{sort}}
	\mathcal{A}_c^{h}(t) = \text{a-sorted}({a_{c,n}(t)}), \forall  n \in \mathcal{N} .
\end{equation}
The WDs in the feasible solutions of the offloading decisions are powered by HAPs with low energy provision cost, and satisfy (\ref{Const:Ddemand}). Based on (\ref{sort}), we obtain the feasible solutions of the offloading decisions.

\item  Reward function: The reward function measures the amount of the energy provision by HAPs at time slot $t$
as
\begin{equation}
	r^{h}(t) =  - \sum_{m=1}^{M} {E_{{h},m}(t)} - \omega_{d},
\end{equation}
where $\omega_{d}$ denotes the penalty of dissatisfying (\ref{Const:Ddemand}). The high-level agent maximizes a series of rewards $r^h$ as
\begin{equation}\label{rh}
	r^{h}=  \sum_{t=0}^{\mathcal{T}-1} \gamma^{h} ~ r^{h}(t).
\end{equation}

\end{itemize}

\subsubsection{Architecture Design of the High-level Agent}
DDPG is employed by the high-level agent.
As the single-agent actor-critic architecture, DDPG uses deep neural networks (DNNs) as a state-action value function   $Q_{\pi^{h}}(\boldsymbol{s}^{h}, \boldsymbol{a}^{h}| \boldsymbol{\theta}^{h} )$ to learn the optimal policy in multi-dimensional continuous action space. Let $\boldsymbol{\theta}^{h}$ denote the parameters of DNNs, and $\boldsymbol{\theta}^{h*}$ denote the optimal policy parameters. There are four DNNs
in DDPG as follows.

The actor network of the high-level agent with parameters $\boldsymbol{\theta}^{h}_{a}$ outputs policy  $\pi^{h}(\boldsymbol{s}^{h} | \boldsymbol{\theta}^{h}_{a})$. Under policy $\pi^{h}(\boldsymbol{s}^{h} | \boldsymbol{\theta}^{h}_{a})$, the high-level agent adopts state $\mathcal{S}^{h}$ as the input of the actor network to output corresponding action $\mathcal{A}^{h}$.

The critic network of the high-level agent with parameters $\boldsymbol{\theta}^{h}_{c}$ evaluates policy  $\pi^{h}(\boldsymbol{s}^{h} | \boldsymbol{\theta}^{h}_{a})$ generated by the actor network  through $Q_{\pi^{h}}(\boldsymbol{s}^{h}, \boldsymbol{a}^{h}| \boldsymbol{\theta}^{h} )$.

The target actor network and target critic network of the high-level agent are used to improve the learning stability. The structure of the target actor network with parameters $\boldsymbol{\theta}_{a,t}^{h}$ is the same as that of the actor network.  The structure of the target critic network with parameters $\boldsymbol{\theta}_{c,t}^{h}$ is the same as that of the critic network.

The input of the DNNs is state $\boldsymbol{s}^{h}$. Motivated by reward $r^{h}$ in (\ref{rh}), the high-level agent learns the optimal policy $\pi^{h*}$ with $\boldsymbol{\theta}^{h*}$ through exploration and training. With obtained optimal policy $\pi^{h*}$, the high-level agent chooses the action that maximizes $Q_{\pi^{h*}}(\boldsymbol{s}^{h}, \boldsymbol{a}^{h}| \boldsymbol{\theta}^{h*})$ for each state.

\subsubsection{Training Process of the High-level Agent}
{The server of the cloud center assists in training the critic network and actor network of the high-level agent.}
 The experience replay buffer stores the experiences of each time slot. The high-level agent randomly selects $k$-size samples of the experiences, and calculates the loss function of the critic network at time slot $t$ as
\begin{equation}
	\begin{split}
		\mathcal{L}^h_{c}(\boldsymbol{\theta}^{h}_c(t)) = \frac{1}{k}\sum_{i=1}^{k}[Q_i(\boldsymbol{s}^{h}_i(t),\boldsymbol{a}^{h}_i(t)|\boldsymbol{\theta}^{h}_c(t)) -
	  (r^{h}_i + \\\gamma^{h} Q_i(\boldsymbol{s}^{h}_i(t+1),\boldsymbol{a}^{h}_i(t+1)|\boldsymbol{\theta}^{h}_c(t)) )]^2.
	\end{split}
  \end{equation}
  By the gradient descent method, the parameters of the critic network is updated as
\begin{equation}\label{xi_1}
  \boldsymbol{\theta}^{h}_c(t+1) \longleftarrow \boldsymbol{\theta}^{h}_c(t)-\beta^{h}_c\bigtriangledown_{\boldsymbol{\theta}^{h}_c(t)}\mathcal{L}^h_{c}(\boldsymbol{\theta}^{h}_c(t)),
\end{equation}
where $\beta^{h}_c$ denotes the learning rate of the critic network.

The action is generated by the actor network of the high-level agent as
\begin{equation}\label{noise}
	\boldsymbol{a}^{h}(t) = \pi^{h}(\boldsymbol{s}^{h}(t)|\boldsymbol{\theta}^{h}_a(t)) + \mathcal{G}(t),
\end{equation}
where $\mathcal{G}(t)$ denotes the Gaussian noise at time slot $t$. The Gaussian noise improves the stability and convergence of the actor network \cite{LRW2023IoT}.

The actor network of the high-level agent aims to maximize the state-action value of the critic network by optimizing the loss function of the actor network as
\begin{equation}
  \mathcal{L}^h_{a}(\boldsymbol{\theta}^{h}_a(t))=\frac{1}{k}\sum_{i=1}^{k} Q(\boldsymbol{s}^{h}(t),\boldsymbol{a}^{h}(t)|\boldsymbol{\theta}^{h}_c(t)).
\end{equation}
To achieve the maximum state-action value of the critic network, the gradient ascent method is adopted to update parameters $\boldsymbol{\theta}^h_a(t)$ as
\begin{equation}\label{xi_2}
  \boldsymbol{\theta}^{h}_a(t + 1) \longleftarrow \boldsymbol{\theta}^{h}_a(t)+\beta^{h}_{a}\bigtriangledown_a \mathcal{L}^h_{a}(\boldsymbol{a}^{h}(t))\bigtriangledown_{\boldsymbol{\theta}^{h}_a(t)}\boldsymbol{a}^{h}(t),
\end{equation}
where $\beta^{h}_{a}$ denotes the learning rate of the actor network.

The target actor network and target critic network of the high-level agent update parameters through soft updating as
\begin{align}\label{xi_3}
	\boldsymbol{\theta}_{a,t}^{h}(t + 1) & = v_a \boldsymbol{\theta}^{h}_a(t) + (1 - v_a) \boldsymbol{\theta}_{a,t}^{h}(t), \nonumber \\
	\boldsymbol{\theta}_{c,t}^{h}(t + 1) & = v_c \boldsymbol{\theta}^{h}_c(t) + (1 - v_c) \boldsymbol{\theta}_{c,t}^{h}(t),
\end{align}
where $0<v_a\le 1$ and $0<v_c\le 1$ are soft updating factors.

\vspace{6mm} 

\subsection{ODO Subproblem}

We formulate the ODO subproblem as a decentralized partially observable Markov decision process (Dec-POMDP), represented by a tuple of low-level agents $(\mathcal{S}^{l}, \mathcal{O}^{l}, \mathcal{A}^{l}, \mathcal{R}^{l},$ Pr$, \gamma^{l} )$, where $\mathcal{S}^{l}$, $\mathcal{O}^{l}$, $\mathcal{A}^{l}$, $\mathcal{R}^{l}$, Pr, and $\gamma^{l}$ represent states, local observations, actions, rewards, the transition probability, and the discount factor, respectively. As low-level agents, WDs adapt actions to maximize the rewards based on local observations. At the beginning of time slot $0$, the global state of the ODO subproblem is initialized as $\boldsymbol{s}^{l}(0)$.  At time slot $t$, WD$_n$ has observation $\boldsymbol{o}^{l}_{n}(t)$ from state $\boldsymbol{s}^{l}(t)$, and then outputs action $\boldsymbol{a}^{l}_{n}(t)$.
 The environment receives the set of $N$ WDs' actions $\boldsymbol{a}^{l}(t) = \{{\boldsymbol{a}^{l}_{n}(t)}\}^N_{n=1}$, and calculates reward $r^{l}_{n}(t) = \mathcal{R}^{l}_n(\boldsymbol{s}^{l}(t), \boldsymbol{a}^{l}(t)), n\in \mathcal{N}$ for $N$ WDs. Then the environment transites to state $\boldsymbol{s}^{l}(t+1)$ according to transition probability Pr$(\boldsymbol{s}^{l}(t+1)|\boldsymbol{s}^{l}(t), \boldsymbol{a}^{l}(t))$.

\begin{itemize}
	\item State space: The global state of the ODO subproblem at time slot $t$ is defined as
	\begin{equation}
		\begin{split}
			\boldsymbol{s}^{l}(t) \!=\! \!\{ \boldsymbol{s}_{p,1}^{l}(t),..., \boldsymbol{s}_{p,M}^{l}(t), &\boldsymbol{s}_{w,1}^{l}(t),..., \boldsymbol{s}_{w,N}^{l}(t),\\ &\boldsymbol{s}_{c,1}^{l}(t),..., \boldsymbol{s}_{c,M}^{l}(t)\},
		\end{split}
	\end{equation}
	where $ \boldsymbol{s}_{p,m}^{l}(t) = \{E^{\text{tot}}_{{h},m}(t),  T-\alpha(t)\}$ represents the state information of HAP$_m$, $\boldsymbol{s}_{w,n}^{l} = \{D_{n}(t), E_{n}(t), a_{c,n}(t)\}$ represents that of WD$_n$, and $\boldsymbol{s}_{c,m}^{l}(t)=\{{h}_{n,m}(t)| n\in \mathcal{N} \}$ represents the channel gains between $N$ WDs and HAP$_m$.
	\item Observation space: $\boldsymbol{o}^{l}_{n}(t)$ denotes the observable state by WD$_n$ from the global state of ODO subproblem $\boldsymbol{s}^{l}(t)$ as
	
	\begin{equation}{\label{On}}
		\begin{split}
			\boldsymbol{o}^{l}_{n}(t) \!=\! \!\{ \boldsymbol{o}_{p,1}^{l}(t),..., \boldsymbol{o}_{p,M}^{l}(t), &\boldsymbol{o}_{w,1}^{l}(t),...,     \boldsymbol{o}_{w,N}^{l}(t),\\ &\boldsymbol{o}_{c,1}^{l}(t),..., \boldsymbol{o}_{c,M}^{l}(t)\},
		\end{split}
	\end{equation}
	where $ \boldsymbol{o}_{p,m}^{l}(t) = \{E^{\text{tot}}_{{h},m}(t),  T-\alpha(t)\}$ represents the observation information of HAP$_m$, $\boldsymbol{o}_{w,n}^{l} = \{D_{n}(t), E_{n}(t), a_{c,n}(t)\}$ represents the observation information of WD$_n$, and $\boldsymbol{o}_{c,m}^{l}(t)=\{{h}_{n,m}(t)| n\in \mathcal{N} \}$ represents the channel gains between $N$ observable
	WDs and HAP$_m$. The size of $\boldsymbol{o}^{l}_{n}(t)$ for WD$_n$ is the same as that of $\boldsymbol{s}^{l}(t)$.
 	If HAP$_m$ is located outside the transmission zone of WD$_n$, the observation information of HAP$_m$ is considered as $\boldsymbol{o}_{p,m}^{l}(t) = \{0,0\}$, and the channel gain between WD$_n$ and HAP$_m$ is considered as $h_{n,m}(t) = 0$ in $\boldsymbol{o}_{c,m}^{l}(t)$. As WD$_n$ can not capture the observation
	 information of the other WDs, the corresponding observation information of other WDs is considered as $\boldsymbol{o}_{w,k}^{l}(t) = \{0,0,0\}$ for $k \in  \mathcal{N} \  \& \ k \ne n$ in (\ref{On}).

\item Action space:
$\mathcal{A}_{n}^{l}(t) = \{x_{n}(t)\}$ represents the action space of WD$_n$ at time slot $t$. $x_{n}(t) = m$, $m \in \mathcal{M}$ represents that WD$_n$ in edge computing mode offloads the computation data to HAP$_m$ at time slot $t$, and $x_{n}(t) = 0$ represents that WD$_n$ in local computing mode processes the computation data locally at time slot $t$.

\item Reward function: In view of observation $\boldsymbol{o}^{l}_{n}(t)$, WD$_n$ outputs offloading policy $\boldsymbol{a}^{l}_{n}(t)$ to interact with the environment, and receives reward $r^{l}_{n}(t)$ as
\begin{equation}\label{rl}
	{r^{l}_{n}(t)} =
	\begin{cases}
	u-(a_{c,n}(t)E_{o,n}(t)+ \!\!\!\!\!\!\!&e_m D_n(t)), \ \ \ \ \\ \ \ \  \ \ \ \ \ \ \ \ \ \ \ \ \ \
	{\text{if}}\ n \in \!\!\!\!&\mathcal{N}_{m}(t),  m \in \mathcal{M};\\
	u-a_{c,n}(t)E_{l,n}(t),&{\text{if}}\ n \in \mathcal{N}_{0}(t);\\
	0,&\text{otherwise},
	\end{cases}
\end{equation}
where $u$ denotes a constant for the non-negative reward. 
The value of $u$ needs to satisfy two requirements: 1) $u >\max\{a_{c,n}(t)E_{o,n}(t)+ e_m D_n(t), a_{c,n}(t)E_{l,n}(t)\}$ for $t \in [0, \mathcal{T}-1]$; 2) the value of $u$ and those of $a_{c,n}(t)E_{o,n}(t)+ e_m D_n(t)$ and $a_{c,n}(t)E_{l,n}(t)$ are similar in the order of magnitude\footnote{We set the value of $u$ as the average upper bound of the energy provision by HAPs for the WP-MEC network with a single WD in edge computing mode over all time slots. Namely, when HAPs with the maximum transmit power $P_{\max}$ provide energy for the WD in edge computing mode during $\mathcal{T}$ time slots,
the average value over all time slots, including the energy consumed by HAPs for broadcasting RF signals and that for processing the received computation data from the WD, is defined as 
\begin{equation}{\label{Psi2}}
	  \begin{split}
		\overline{\Psi} = \frac{\mathcal{T} M  P_{\max} T +  \mathop{max}\limits_{m}\{e_m\} \sum_{t=1}^{\mathcal{T}}    \mathbb{E}\left(D_n(t)\right) }{\mathcal{T} },
	  \end{split}
  \end{equation}
where $\mathbb{E}\left(D_n(t)\right)$ represents the expectation of offloaded bits from WD$_n$ at time slot $t$.}.
{The TMADO framework aims to maximize the cumulative sum of (\ref{rl}), i.e., the total reward of low-level agents, by minimizing the energy provision by HAPs in (\ref{Objective:AP}). In the solution of $\mathbf{P0}$, HAPs only need to provide the required amount of energy for WDs in local computing mode or edge computing mode in  (\ref{Objective:AP}).}

Based on (\ref{sh}) and (\ref{sort}), HAPs determine the feasible solutions of the offloading decisions $(\mathcal{N}- \mathcal{N}_{\text{-1}}(t)) $, and broadcast them at the beginning of time slot $t$. Then WDs receive the feasible solutions of the offloading decisions, and output actions $\boldsymbol{a}^{l}(t)$ based on local observations.
For WD$_n$, $n\in \mathcal{N}$, the ODO subproblem aims to find the optimal policy $\pi^{l*}_{\boldsymbol{\theta}^{l}_{a, n}}$ that maximizes the long-term accumulated discounted reward as

\begin{equation}
	\label{P1}
			~~\underset{\pi^{l*}_{\boldsymbol{\theta}^{l}_{a, n}}}    {\text{max}}~~~
			\mathbb{E}\left[\sum_{t=0}^{\mathcal{T}-1} \gamma^{l}_t ~ r^{l}_{n}(t)\right].
\end{equation}
Let $\boldsymbol{\theta}^{l}_{a, n}$ denote the parameters of the actor network of low-level agent $n$,

\end{itemize}

\subsubsection{Architecture Design of Low-level Agents}
For the architecture design of low-level agents, IPPO is employed by low-level agents. IPPO, consisting of the actor networks and critic networks, is the multi-agent actor-critic architecture.
The number of the actor networks of low-level agents is the same as that of WDs. The actor network of low-level agent $n$ with parameters $\boldsymbol{\theta}^{l}_{a, n}, n \in \mathcal{N}$ outputs policy $\pi_{\boldsymbol{\theta}^{l}_{a, n}}^{l}(\boldsymbol{a}^{l}_n(t) | \boldsymbol{o}^{l}_n(t))$, which is predicted distribution of action $\boldsymbol{a}^{l}_n(t)$ given local observation $\boldsymbol{o}^{l}_n(t)$. By adding a softmax function \cite{HZW2023TMC}, the actor network of low-level agent $n$ outputs transition probability Pr$_i(\boldsymbol{o}^{l}_n(t))$ of WD$_n$'s actions to provide the discrete action, which follows the categorical distribution as
\begin{equation}
	\begin{aligned}
		\!\pi^{l}_{\boldsymbol{\theta}^{l}_{a, n}} \left(\boldsymbol{a}^{l}_n(t) \mid  \boldsymbol{o}^{l}_n(t)\right)\! = \! \prod_{i=1}^{M+1}\! \text{Pr}_i & \left(\boldsymbol{o}^{l}_n(t)\right) \!I_{\left\{\boldsymbol{a}^{l}_n(t)=i\right\}},
	\end{aligned}
\end{equation}
and
\begin{equation}
	\begin{aligned}
		\! \sum_{i=1}^{M+1}\! \text{Pr}_i\left(\boldsymbol{o}^{l}_n(t)\right)\!=\!1,
	\end{aligned}
\end{equation}
where $M+1$ represents the number of WD$_n$'s actions in the action space $\mathcal{A}_{n}^{l}(t)$.

Besides, the number of the critic networks of low-level agents is the same as that of WDs. The critic network of low-level agent $n$ with parameters $\boldsymbol{\theta}^{l}_{c, n}, n \in \mathcal{N}$ evaluates state value function $V_n(t), n \in \mathcal{N}$ in (\ref{vn}),
and guides the update of the actor network during the training process.

\subsubsection{Training Process of Low-level Agents}
For the training and execution of low-level agents, centralized training and decentralized execution (CTDE) mechanisms are employed by low-level agents.
The CTDE mechanisms do well in tackling the non-stationary issue in multi-agent DRL algorithms by reducing the impact of the interference from irrelevant state information of other agents,
and show promising performance in distributed scenarios \cite{Hairi2022FiniteTimeCA}.
 To employ CTDE mechanisms, we consider that the server of the cloud center assists in training the critic networks and actor networks of low-level agents in a centralized manner \cite{GZ2023TMC}, and the trained low-level agents determine actions in a distributed manner. During centralized training, the cloud computing layer has fully-observable access to the states, actions, and rewards of other low-level agents. During decentralized execution, the low-level agent has partially-observable access to the states, actions, and rewards of them, i.e., the low-level agent relies on its local observations to output action for the computation offloading, without the states, actions, and rewards of other WDs.

We introduce the critic networks and actor networks of low-level agents in detail. The critic networks estimate the unknown state value functions to generate update rules for the actor networks. The actor networks output policies $\pi_{\boldsymbol{\theta}^{l}_{a, n}}^{l}(\boldsymbol{a}^{l}_n(t) | \boldsymbol{o}^{l}_n(t)), n \in \mathcal{N}$ to maximize the fitted state values.

The objective function for the critic network of low-level agent $n$ at time slot $t$, i.e., the state value function, is formulated as
\begin{equation}\label{vn}
	V_{n}(t) =V_n\left(\boldsymbol{o}^{l}_{n}(t)\right)=\mathbb{E}\left[\sum_{t^{\prime}=t}^{\mathcal{T}-1} \gamma^{l} r^{l}_{n}(t^{\prime})\right],
  \end{equation}
  where $ \gamma^{l} \in [0, 1]$ is the discount factor.
The advantage function of low-level agent $n$ at time slot $t$, i.e., comparing the old
policy and the current policy in terms of variance, is formulated as
\begin{equation}\label{af}
	\begin{split}
	A_{n}(t) &=A_n \left(\boldsymbol{o}^{l}_{n}(t), \boldsymbol{a}^{l}_{n}(t)\right)
	\\  &=   r^{l}_{n}(t)+\gamma^{l} V_{n}(t+1)-V_{n}(t).
	\end{split}
\end{equation}
The state value function $V_n\left(\boldsymbol{o}^{l}_{n}(t)\right)$ in (\ref{vn}) is updated by the mean squared error loss function as
\begin{equation}
\mathcal{L}^l_{n}(t)=\mathbb{E}\left[r^{l}_n(t)+\gamma^{l} V_n\left(\boldsymbol{o}^{l}_n(t+1)\right)-V_n\left(\boldsymbol{o}^{l}_n(t)\right)\right]^2.
\end{equation}

Based on (\ref{af}), the objective function of the actor network of low-level agent $n$ at time slot $t$, i.e., the surrogate objective of low-level agent $n$, is formulated as
\begin{equation}\label{surrogate}
J^l\!\!\left(\!\pi^{l}_{\boldsymbol{\theta}^{l}_{a, n}}\!\right)\!\!=\!\mathbb{E}_{(\boldsymbol{s}^{l}, \boldsymbol{a}^{l}\!)} \min \left(\varrho(t) A_n(t), \operatorname{clip}(\varrho, 1\!-\!\epsilon, 1\!+\!\epsilon) A_n(t)\right)\!,
\end{equation}
where $\varrho(t)=\frac{\pi^{l}_{\boldsymbol{\theta}^{l}_{a, n}}\left(\boldsymbol{a}_n(t) \mid \boldsymbol{o}_n(t)\right)}{\pi^{l}_{\boldsymbol{\theta}^{l,old}_{a, n}}\left(\boldsymbol{a}_n (t)\mid \boldsymbol{o}_n (t)\right)}$ represents the truncated importance sampling factor, $\pi^{l}_{\boldsymbol{\theta}^{l}_{a, n}}\left(\boldsymbol{a}^{l}_n(t) \mid \boldsymbol{o}^{l}_n(t)\right)$ represents the current policy, $\pi^{l}_{\boldsymbol{\theta}^{l,old}_{a, n}}\left(\boldsymbol{a}^{l}_n(t) \mid \boldsymbol{o}^{l}_n(t)\right)$ represents the old policy,
the clip function makes $r^{l}_n(t)$ in the value range $[1-\epsilon, 1+\epsilon]$ to train the actor network of low-level agent $n$,
and $\epsilon$ represents a positive hyperparameter.
To evaluate the difference between the old and current policies, the surrogate objective of low-level agent $n$ uses the importance sampling strategy \cite{{HZW2023TMC}} to treat the samples from the old policy as the surrogate of new samples in training the actor network of low-level agent $n$.

\subsection{RO Subproblem}
With given \{$\boldsymbol{{P}_{{h}}(t)}$, $\alpha(t)$, $\mathcal{N}_\text{-1}(t)$, $\boldsymbol{\mathcal{X}(t)}$\}, we analyze the RO subproblem as follows. We first consider the case when WD$_n$ chooses local computing mode at time slot $t$, i.e.,  $ n \in \mathcal{N}_0(t) $.


\begin{lemma}{\label{Lemma2}}
With given {\rm $\boldsymbol{{P}_{{h}}(t)}$, $\alpha(t)$, $\boldsymbol{\mathcal{X}(t)}$}, the optimal computing delay {\rm $ \tau_{l,n}(t), n \in \mathcal{N}_0(t)$} and the CPU frequency {\rm $ f_{n}(t)$} of WD{\rm $_{n}$} in local computing mode at time slot {\rm $t$} are respectively given as
\begin{equation}{\label{close1}}
	\tau_{l,n}^*(t) = T,
\end{equation}
and
\begin{equation}{\label{close2}}
    f_{n}^*(t) = \frac{{C}_{n}{D}_{n}(t)}{T}.
\end{equation}
\end{lemma}

\begin{IEEEproof}
	With given $\boldsymbol{\mathcal{X}(t)}$, based on (\ref{Eini}), the initial energy in WD$_n$ at time slot $t+1$ is determined by the initial energy in WD$_n$ and the energy consumption of WD$_n$ for local computing at time slot $t$. With given $\boldsymbol{{P}_{{h}}(t)}$, $\alpha(t)$,
	the amount of the initial energy in WD$_n$ at time slot $t+1$ decreases with the amount of the energy consumed by WD$_n$ for local computing at time slot $t$,
	which is given as follows based on (\ref{tloc1}) and (\ref{Eloc1}).
	\begin{equation}{\label{x}}
		E_{l,n}(t)= k_n C_{n} D_{n}(t) f^2_{n}(t) = \frac{k_n( C_{n}D_{n}(t))^3}{\tau_{l,n}(t)^2}.
	\end{equation}
	  $E_{l,n}(t)$ decreases with $\tau_{l,n}(t)$.
	 Based on (\ref{tloc1}) and the monotonicity of $E_{l,n}(t)$ with respect to $\tau_{l,n}(t)$, we have (\ref{close2}).
\end{IEEEproof}
 Then we consider the case when WD$_n$ chooses edge computing mode at time slot $t$, i.e.,  $ n \in \mathcal{N}_m(t) $.

\begin{lemma}{\label{Lemma3}}
	With given $\boldsymbol{{P}_{{h}}(t)}$, $\alpha(t)$, $\boldsymbol{\mathcal{X}(t)}$, the optimal duration that WD$_n$ offloads the computation data to HAP$_m$ {\rm $ \tau_{o,n,m}(t)$, $n \in \mathcal{N}_m(t)$, $m \in \mathcal{M}$} at time slot {\rm $t$} is
\begin{equation}\label{tau_o_n_m}
	\tau_{o,n,m}^*(t) = \frac{v {D}_{n}(t) }{B {\mathrm{log}}_{2}\left(1+\frac{P_{n} {h}_{n,m}(t)}{{N}_{0}}\right)}.
\end{equation}
\end{lemma}

\begin{IEEEproof}
	 Based on Lemma \ref{Lemma2} and (\ref{Const:OffE}), the amount of the energy consumed by WD$_n$ for edge computing is
	 \begin{equation}\label{close3}
		E_{o,n}(t)= \left(P_{n}+P_{c,n}\right) \tau_{o, n,m}(t).
	\end{equation}
	It is easy to observe that $E_{o,n}(t)$ increases with $\tau_{o,n,m}(t)$.
	 Based on (\ref{tau_{n,m}}), (\ref{Const:OffT_tau}),   (\ref{Const:OffT}), and the monotonicity of $E_{o,n}(t)$ with respect to $\tau_{o,n,m}(t)$,  the minimum duration that WD$_n$ offloads the computation data to HAP$_m$ in (\ref{tau_o_n_m}) ensures that WD$_n$ in edge computing mode  successfully offloads the computation data of $ D_{n}(t) $ bits to HAP$_m$.
\end{IEEEproof}

\begin{algorithm}[t] 
	\small
    \SetAlgoLined 
	\caption{TMADO Framework to Solve $\mathbf{P}_{0}$}
	\KwIn{Parameters of the critic network, the actor network, the target critic network, and the target actor network as $\boldsymbol{\theta}^{h}_{c}$, $\boldsymbol{\theta}^{h}_{a}$, $\boldsymbol{\theta}^{h}_{c,t}$, and $\boldsymbol{\theta}^{h}_{a,t}$,  parameters of the critic networks and actor networks as $\boldsymbol{\theta}^{l}_{c, n}$ and $\boldsymbol{\theta}^{l}_{a, n}$, $n \in \mathcal{N}$, and experience replay buffers.}
	\KwOut{Optimal policy $\pi^{h}$ and $\left\{\pi^{l}_{\boldsymbol{\theta}^{l}_{a, n}}\right\}_{n=1}^N$.}
	Randomly initialize parameters of the critic network, the actor network, the target critic network, and the target actor network as $\boldsymbol{\theta}^{h}_{c}$, $\boldsymbol{\theta}^{h}_{a}$, $\boldsymbol{\theta}^{h}_{c,t}$, and $\boldsymbol{\theta}^{h}_{a,t}$,  parameters of the critic networks and the actor networks as $\boldsymbol{\theta}^{l}_{c, n}$ and $\boldsymbol{\theta}^{l}_{a, n}$, $n \in \mathcal{N}$, and experience replay buffers\;
	\For{$\text{Iter}_1 = 1, 2, \dots, \text{Iter}_{\max}$}{
		Clean experience replay buffers\;
		\For{$t = 0, 1, \dots,\mathcal{T}-1$}{
			Get global state of the WCDO subproblem {$\boldsymbol{s}^{h}(t)$}\;
			The high-level agent outputs action $\boldsymbol{a}^{h}(t) \sim \pi^{h}$\;
			Interact with the environment by $\boldsymbol{a}^{h}(t)$\;
			Distribute $\boldsymbol{a}^{h}(t)$ to low-level agents\;
					\For{$n = 1, 2, \dots,N$}{
					Get local observation {$\boldsymbol{o}^{l}_{n}(t)$}\;
					Low-level agent $n$ outputs action $\boldsymbol{a}^{l}_{n}(t) \sim \pi_{\boldsymbol{\theta}^{l}_{a, n}}^{l}$\;
			Interact with the environment by $\boldsymbol{a}^{l}_{n}(t)$\;
			Compute reward $r^{l}_n(t)$ according to (\ref{rh})\;
			Store the tuple $\left(\boldsymbol{o}^{l}_{n}(t),
			\boldsymbol{a}^{l}_{n}(t), r^{l}_{n}(t)\right)$ and state $\boldsymbol{s}^{l}(t+1)$ in the experience replay buffer of low-level agent $n$\;
			}
		
			Compute reward $r^{h}(t)$ according to (\ref{rl})\;
			Store the tuple $\left(\boldsymbol{s}^{h}(t), \boldsymbol{a}^{h}(t), r^{h}(t), \boldsymbol{s}^{h}(t+1)\right)$ in the experience replay buffer of the high-level agent\;
			Update the critic network, the actor network, the target critic network, and the target actor network of the high-level agent\;
		}
		Store current policy $\pi^{l}_{\boldsymbol{\theta}^{l,old}_{a, n} }\leftarrow \pi_{\boldsymbol{\theta}^{l}_{a, n}}^{l} $ for each low-level agent\;
			\For {$k = 1, 2, \dots,M_1$}{
				\For {$n = 1, 2, \dots,N$}{
					Compute state value function according to (\ref{vn})\;
					Compute advantage function according to (\ref{af})\;
					Update the critic network and the actor network of low-level agent $n$\;
				}
			}
	}
\end{algorithm}
  
\subsection{Proposed TMADO Framework}

Algorithm 1 shows the pseudo-code of the proposed TMADO framework. To be specific, we initialize the actor-critic network parameters of the high-level agent and low-level agents, and the experience
replay buffers of the high-level agent and low-level agents in line 1. Then we start a loop for sampling and training in line 2. The high-level agent outputs action by the DDPG algorithm in lines 5-6. After interacting with the environment in line 7, the actions of the high-level agent $\mathcal{A}^{h}(t)$ in (\ref{A^h}) are broadcasted to low-level agents, and used as the starting point for low-level agents' action exploration in line 8. Low-level agents output actions by the IPPO algorithm in lines 10-11. After interacting with the environment in line 12, low-level agent $n$ stores the sampled experience $\left(\boldsymbol{o}^{l}_n(t), \boldsymbol{a}^{l}_n(t), r^{l}_{n}(t)\right)$ and $\boldsymbol{s}^{l}(t+1)$ into the experience replay buffer of low-level agent $n$ in lines 13-14. The high-level agent stores the sampled experience $\left(\boldsymbol{s}^{h}(t), \boldsymbol{a}^{h}(t), r^{h}(t), \boldsymbol{s}^{h}(t+1)\right)$ into the experience replay buffer of the high-level agent in lines 16-17. Then the critic network, the actor network, the target critic network, and the target actor network of the high-level agent are updated in line 18. We update the critic networks and actor networks of low-level agents for $M_1$ times as sample reuse in PPO \cite{YYX2023ICDE} in lines 20-27. Each time, the experience replay buffer is traversed to conduct mini-batch training. 



\subsection{Computation Complexity Analysis}

We provide the computational complexity analysis of the training process as follows. For the high-level agent, let $L_a$, $L_c$, $n_{l_a}$, $n_{l_c}$, and $T_\text{ep}$ denote the number of layers for the actor network, that for the critic network, the number of neurons in layer $l_a$  of the actor network, the number of neurons in layer $l_c$ of the critic network, and the number of episodes,  respectively. Then the complexity of the high-level agent can be derived as $O( T_\text{ep} \mathcal{T}(
	\sum_{l_a=0}^{L_a-1} n_{l_a}n_{{l_a}+1}+\sum_{l_c=0}^{L_c-1} n_{l_c}n_{{l_c}+1}))$ \cite{LRW2023IoT}. Since low-level agents adopt mini-batches of experience to train policies by maximizing the discounted reward in  (\ref{P1}), based on (\ref{surrogate}), we adopt the sample complexity in \cite{Hairi2022FiniteTimeCA} to characterize the convergence rate by achieving
	\begin{equation}
		\mathbb{E}\Big[\Big\lVert \nabla_(\pi^{l}_{\boldsymbol{\theta}^{l}_{a, n})} J^l\!\left(\pi^{l}_{\boldsymbol{\theta}^{l}_{a, n}}\right) \Big\lVert^2\Big]  \leq \epsilon.
	\end{equation}
	Then the sample complexity of low-level agents is $O(\epsilon^{-2}\log(\epsilon^{-1}))$ \cite{Hairi2022FiniteTimeCA}. The complexity of solving (\ref{close1})-(\ref{tau_o_n_m}) is $O(1)$.
	In summary, the overall computational complexity of the training process is $ O(T_\text{ep} \mathcal{T}(\sum_{l_a=0}^{L_a-1} n_{l_a}n_{{l_a}+1}+\sum_{l_c=0}^{L_c-1} n_{l_c}n_{{l_c}+1}) + \epsilon^{-2}\log(\epsilon^{-1}))$.

\section{Simulation Results}

\begin{table}[t]
	\centering
	\caption{Simulation Parameters}
	\label{tab:1}
	\renewcommand\arraystretch{1.2}
	\setlength{\tabcolsep}{1.0mm}
	\begin{tabular}{|l|l|}
		\hline
		Descriptions & Parameters and values\\
		\hline
		\hline
					  & $M$ = $3$, $N = 10$\\
		  & $\mathcal{T} = 100$, $T = 0.4$ s \\
		Network model   &  $R_t = 25 $ m, $E_b = 100 $ mJ \\
					 & $\lambda = 50$, $D_p = 10^3$ bit \\
					 &  $D_{{th}}=3.5 \times 10^5$ bit\\
					
		\hline
		EH model & $\mu = 0.51$, $P_{\max} = 3$ W\\
		\hline
		Local computing mode & $k_n = 10^{-27}$, $C_n = 10^{3}$ cycle/bit, \\
		
		\cite{WXJ2023DCO}	&  $f_{\max}=0.3$ GHz\\
		\hline
							& $B = 1$ MHz, $N_0 = 10^{-9}$ W \cite{WF2018TWC}\\
							
		Edge computing mode	&   $P_n = 0.1$ W, $P_{c,n} = 10^{-3}$ W\\
							& $e_m = 1 \times 10^{-6}$ J/bit \cite{BT2021TWC}, $v = 1.1$ \\
		\hline
		High-level agent 	     			 &  $\boldsymbol{\theta}^{h}_a = 10^{-5} $, $\boldsymbol{\theta}^{h}_c = 10^{-5} $ \\
		(DDPG)	& $v_a = 10^{-4}$, $v_c = 10^{-4}$, $\gamma^h = 0.95$\\
						 
		\hline
		Low-level agents           & $\boldsymbol{\theta}^{l}_a = 10^{-5} $, $\boldsymbol{\theta}^{l}_c = 10^{-5} $ \\
		(IPPO)				& $\gamma^l = 0.99$, $\epsilon = 0.2$, $u=3.65$\\ 		
		\hline
	\end{tabular}
\end{table}

\begin{figure*}[t]
	\begin{minipage}[t]{0.33\linewidth}
	\centering
	\includegraphics[width=5.755cm]{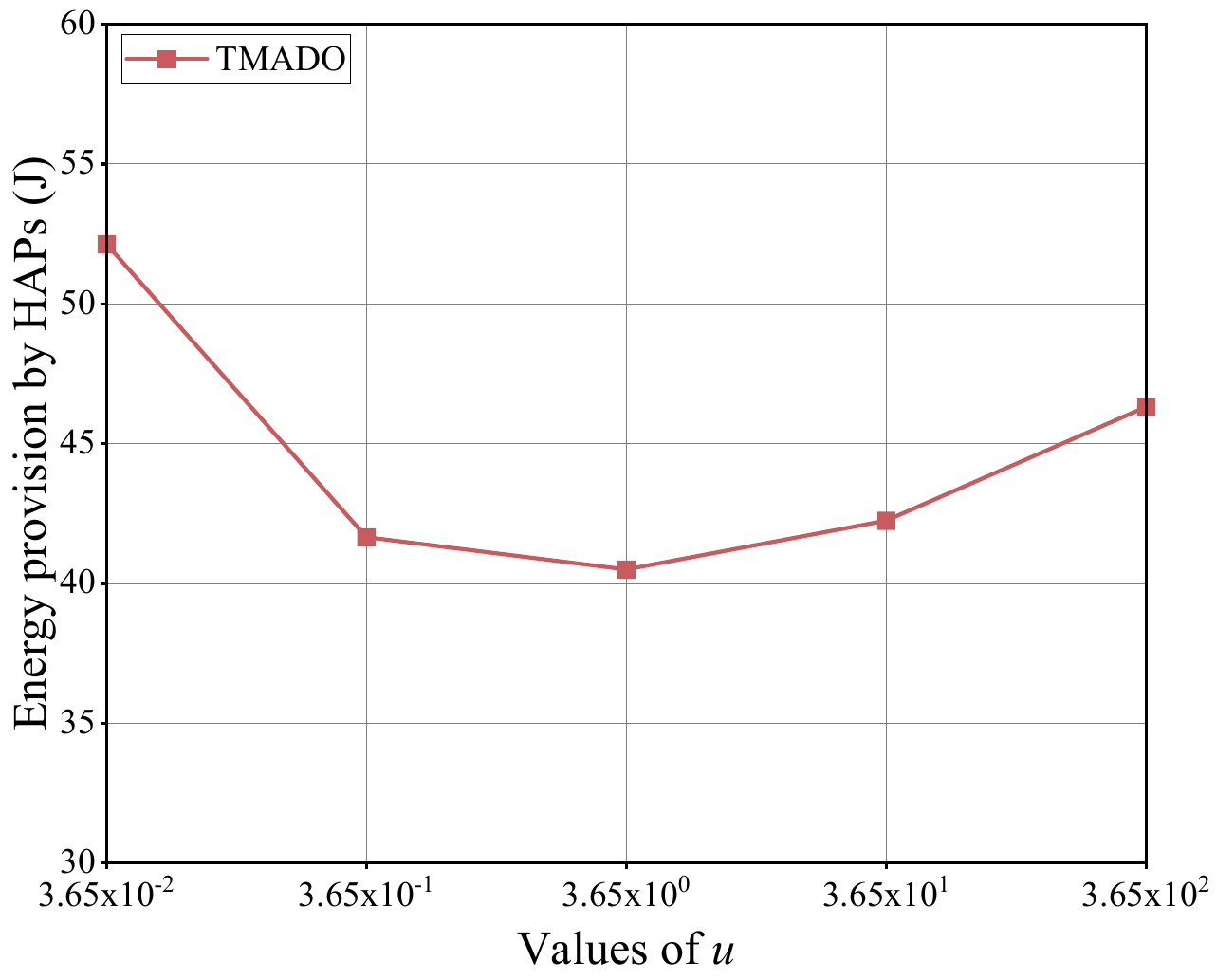}
	\caption{The energy provision (EP) by HAPs \\ versus  $u$.}
	\label{Exp0HyperPara}
	\end{minipage}%
	\begin{minipage}[t]{0.32\linewidth}
	\centering 
	\includegraphics[width=5.87cm]{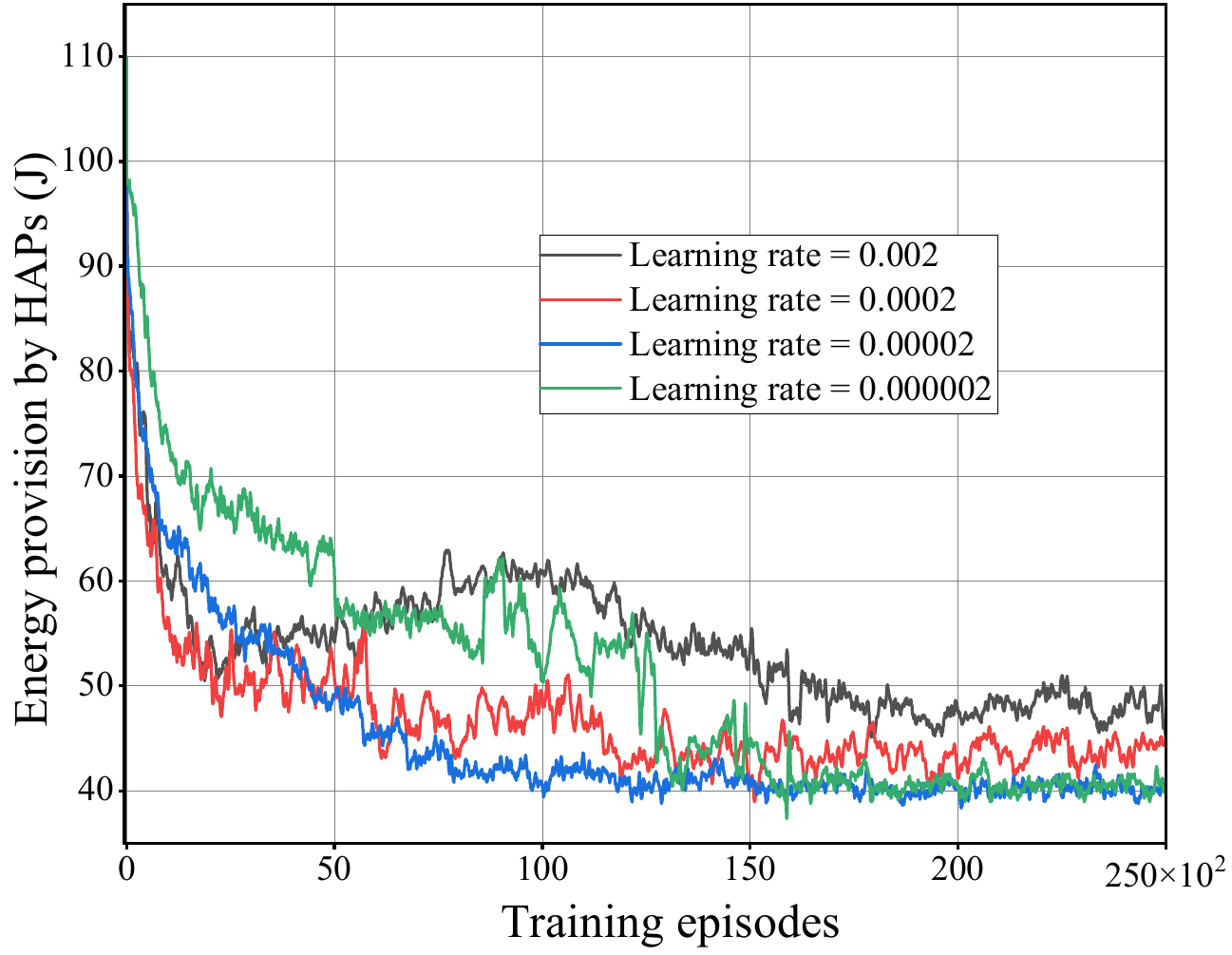}
	\caption{The EP by HAPs under the proposed \\ TMADO scheme versus training episodes.}
	\label{Exp1high}
	\end{minipage}
	\begin{minipage}[t]{0.33\linewidth}
	\centering
	\includegraphics[width=6.1cm]{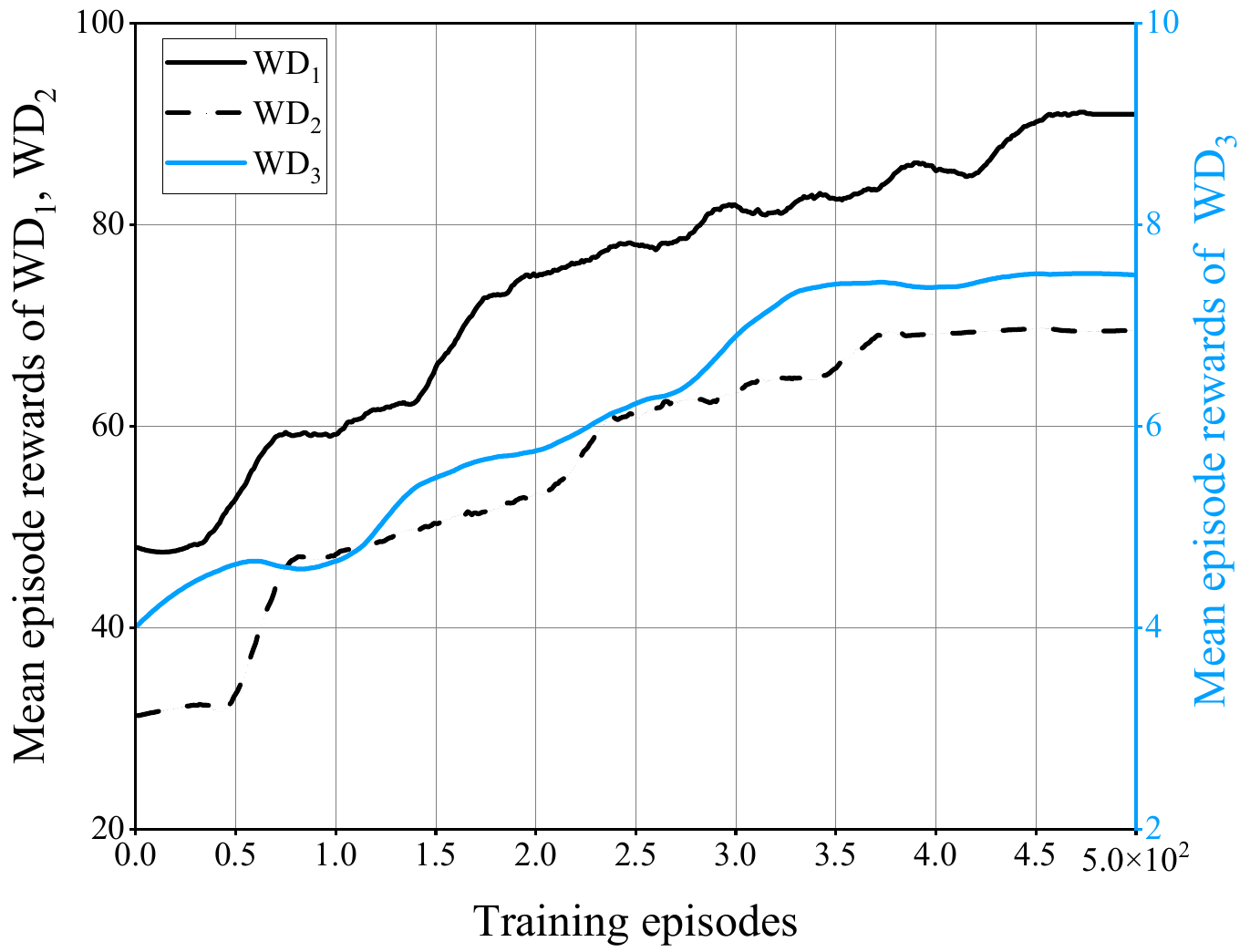}
	\caption{The mean episode rewards of WDs under the proposed TMADO scheme versus training episodes.}
	\label{Exp1low}
	\end{minipage}
\end{figure*}

In this section, we set the WP-MEC network with the area of 100 m $\times$ 100 m field \cite{GZ2023TMC}, where HAPs are uniformly distributed, and WDs are randomly distributed. The distance between WD$_n$ and HAP$_m$  is $d_{n,m}$, and then the large-scale fading component is ${\sigma}_{L,n,m} \!=\! A_d\left(\frac{3 \times 10^8}{4\pi f_{\text{cf}} d_{n,m}}\right)^{d_e}$ \cite{BSZ2018TWC}, where $A_d$ with $4.11$ is the antenna gain, $f_{\text{cf}}$ with 915 MHz is the carrier frequency, and $d_e$ with $2$ is the path loss exponent. The small-scale Rayleigh fading follows an exponential distribution with unit mean. The number of the arrived data packets in WDs follows an independent Poisson point process with rate $\lambda = 50$. Other simulation parameters are given in TABLE \ref{tab:1}.

\begin{figure*}[t]
	\begin{minipage}[t]{0.33\linewidth}
	\centering
	\includegraphics[width=6cm]{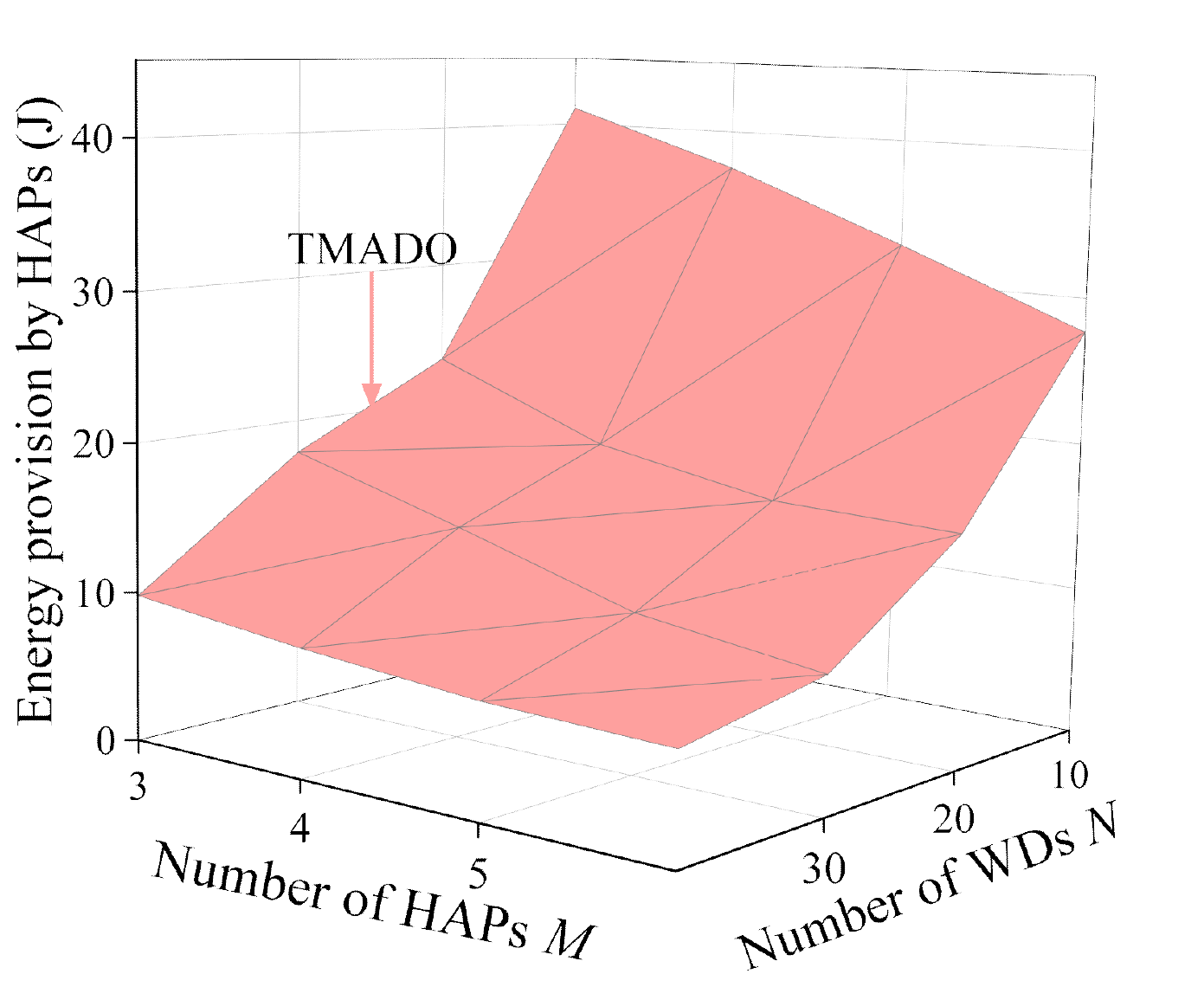}
	\caption{The EP by HAPs under the proposed \\ TMADO scheme versus the  number \\ of HAPs $M$ and that of WDs $N$.}
	\label{Exp3Number}
	\end{minipage}%
	\begin{minipage}[t]{0.33\linewidth}
	\centering 
	\includegraphics[width=6cm]{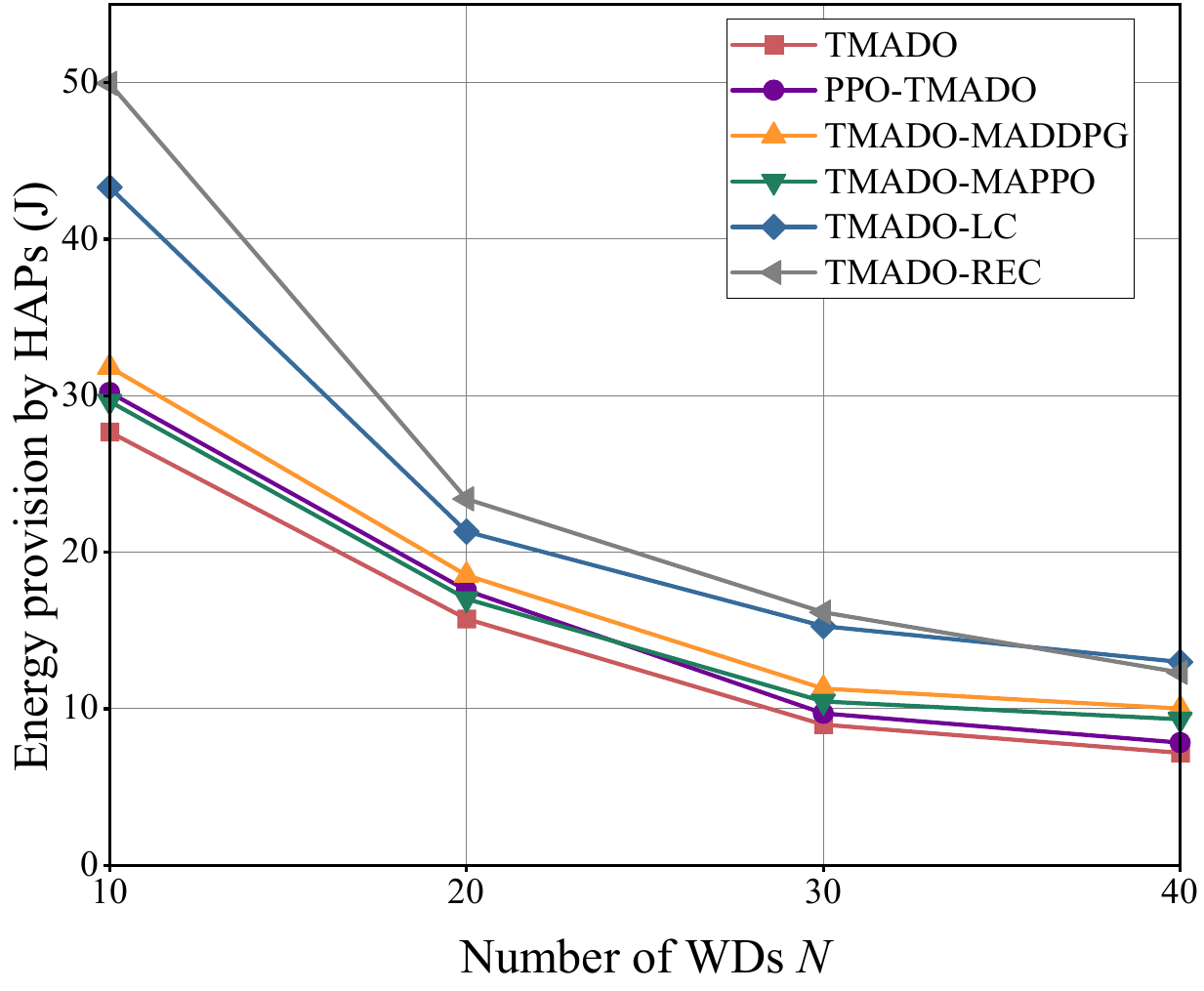}
	\caption{The EP by HAPs under six schemes \\ versus the  number of WDs $N$ with $M=6$.}
	\label{Exp3Number-WD}
	\end{minipage}
	\begin{minipage}[t]{0.33\linewidth}
	\centering
	\includegraphics[width=6cm]{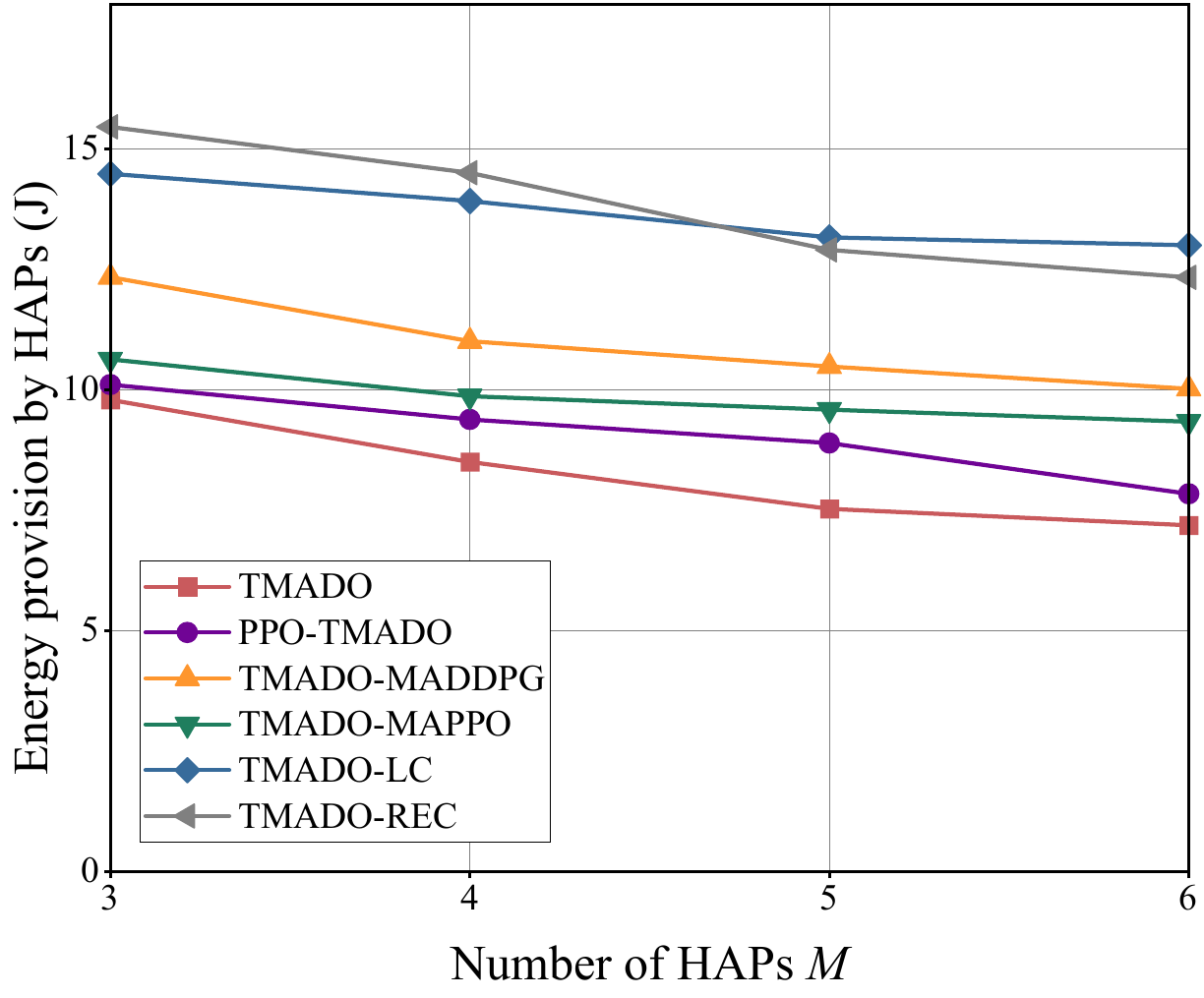}
	\caption{The EP by HAPs under six schemes \\ versus the number of HAPs $M$ with $N=40$.}
	\label{Exp3Number-HAP}
	\end{minipage}
\end{figure*}

\begin{figure*}
	\begin{minipage}[t]{0.33\linewidth}
	\centering
	\includegraphics[width=5.84cm]{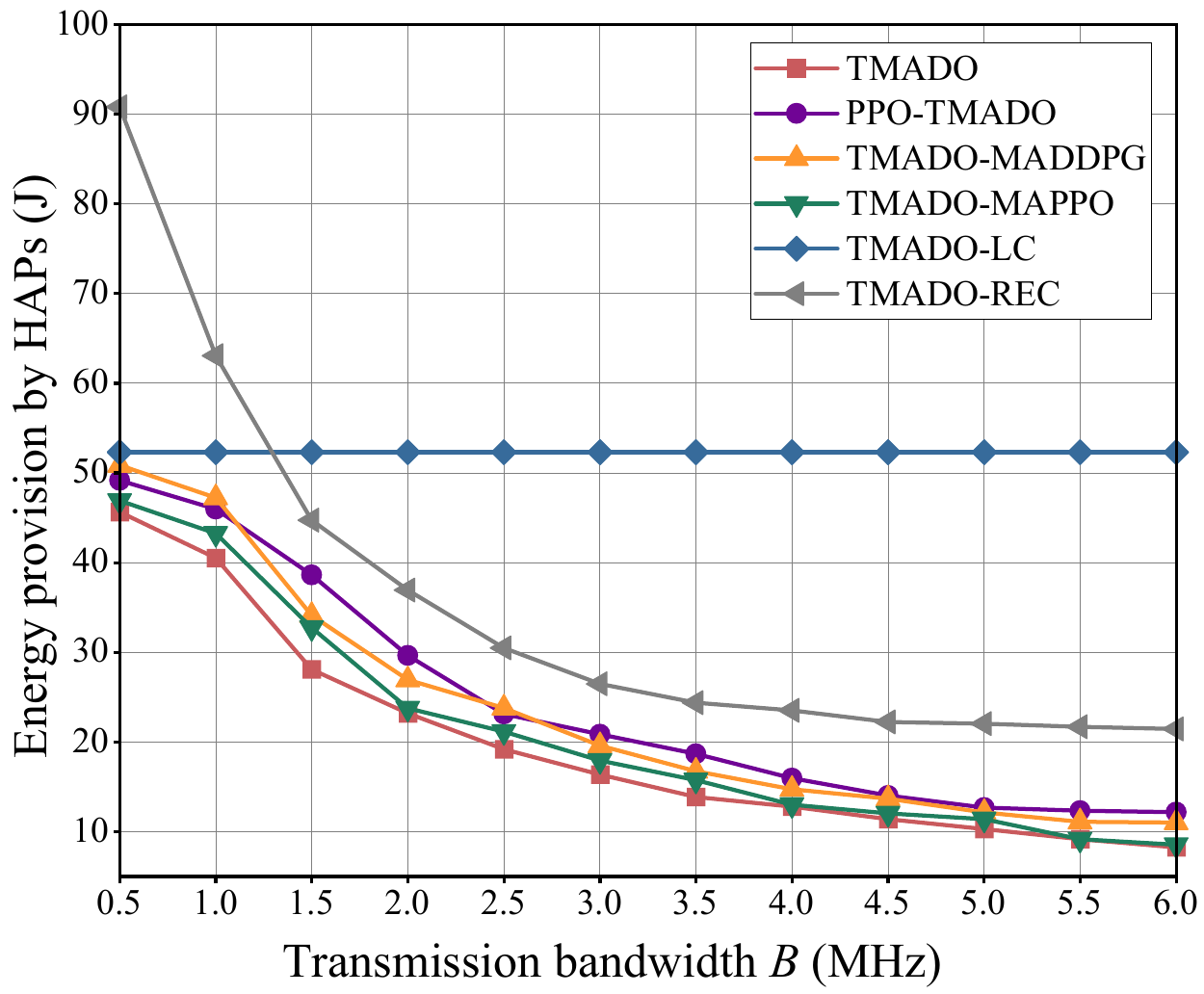}
	\caption{The EP by HAPs under six schemes \\ versus the transmission bandwidth $B$.}
	\label{Exp4Bandwidth}
	\end{minipage}%
	\begin{minipage}[t]{0.33\linewidth}
	\centering
	\includegraphics[width=6cm]{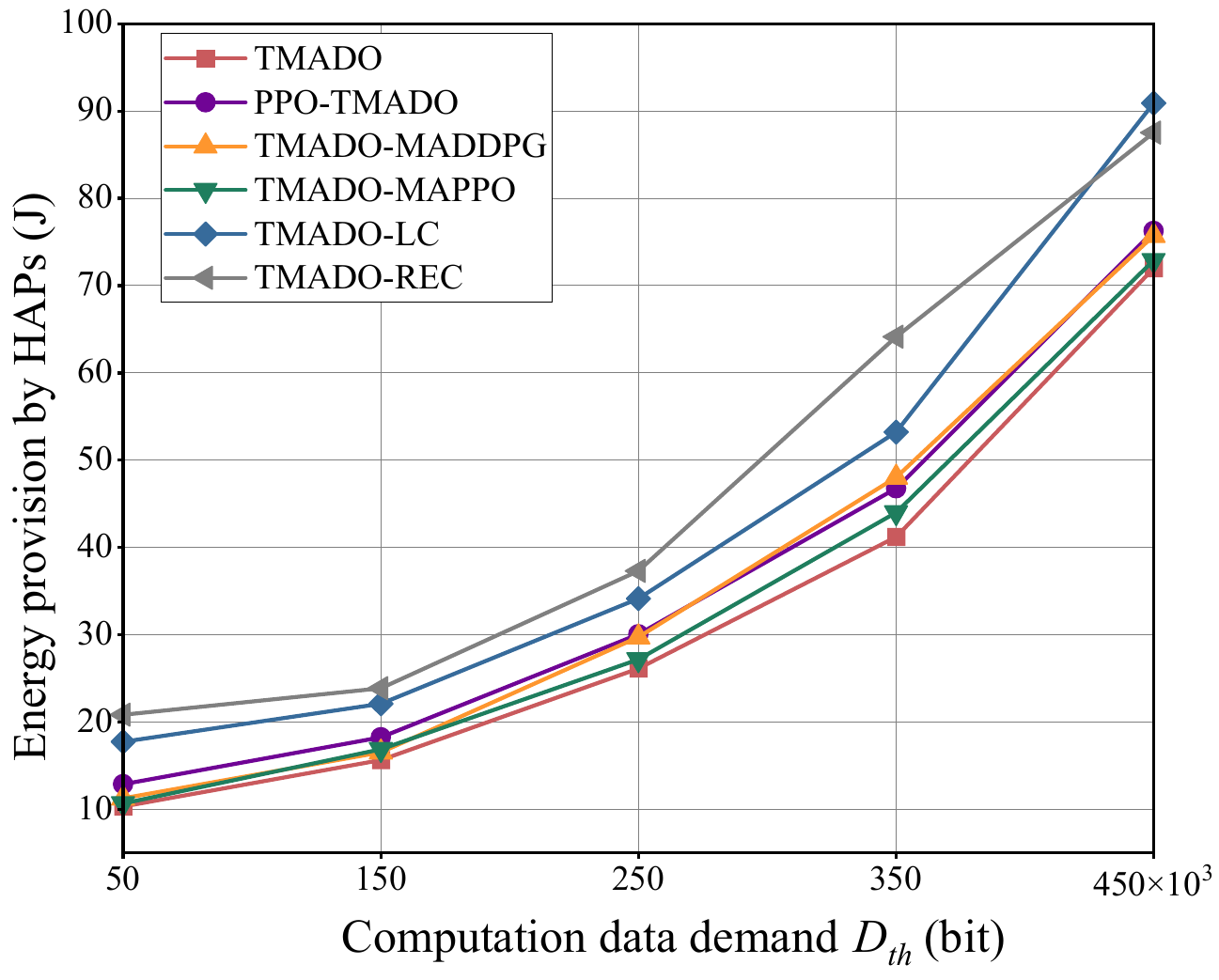}
	\caption{The EP by HAPs under six schemes \\ versus the computation data demand $D_{th}$.}
	\label{Exp7Demand}
	\end{minipage}
	\begin{minipage}[t]{0.33\linewidth}
	\centering
	\includegraphics[width=5.68cm]{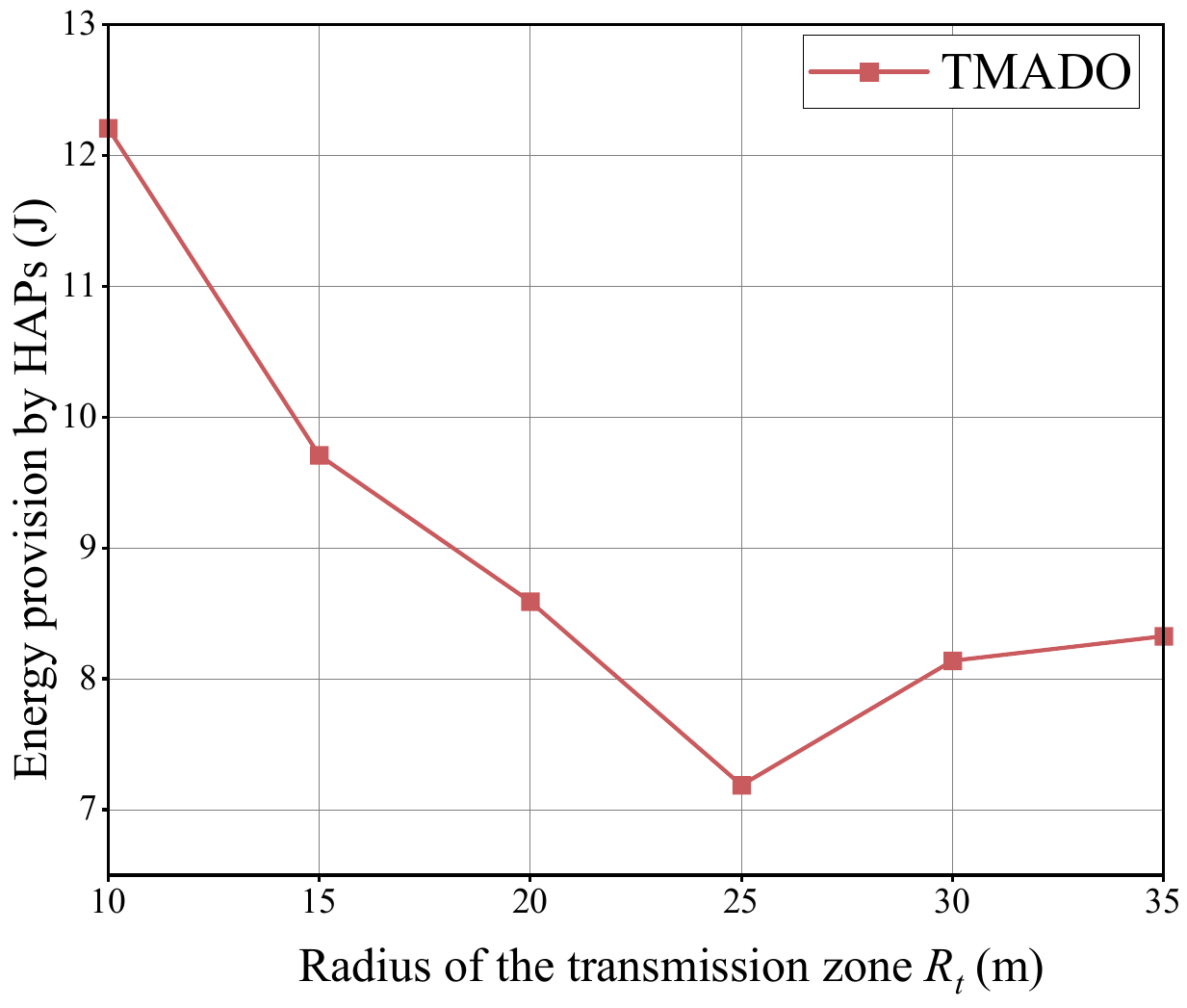}
	\caption{The EP by HAPs under the proposed TMADO scheme versus the radius of the transmission zone $R_{t}$ with $M=6$ and $N=40$.}
	\label{Exp6Possion}
	\end{minipage}
\end{figure*}

\begin{figure}[t]
	\centering
	\subfloat[]{
		\includegraphics[width=4.25cm]{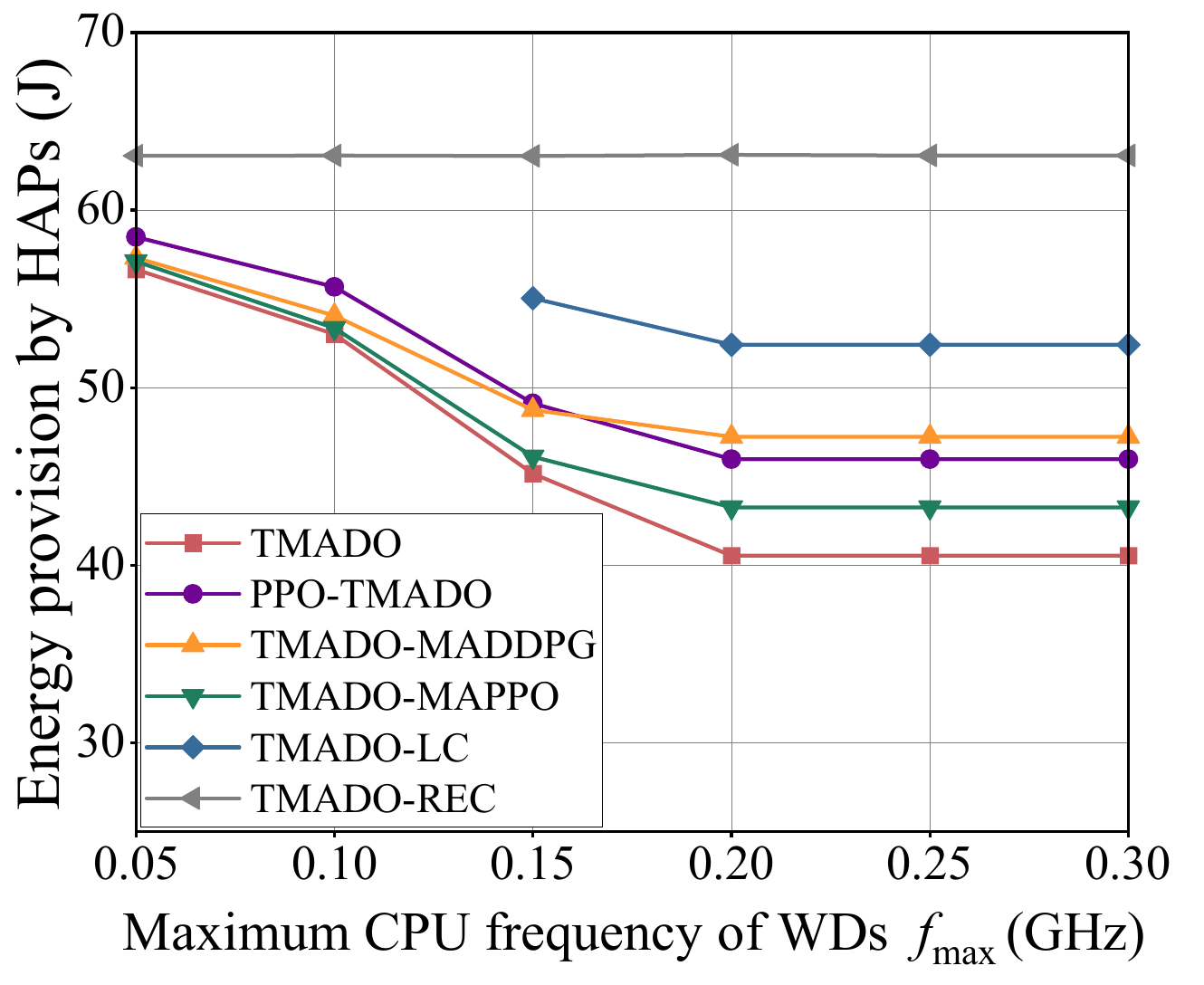}}
	\subfloat[]{
		\includegraphics[width=4.25cm]{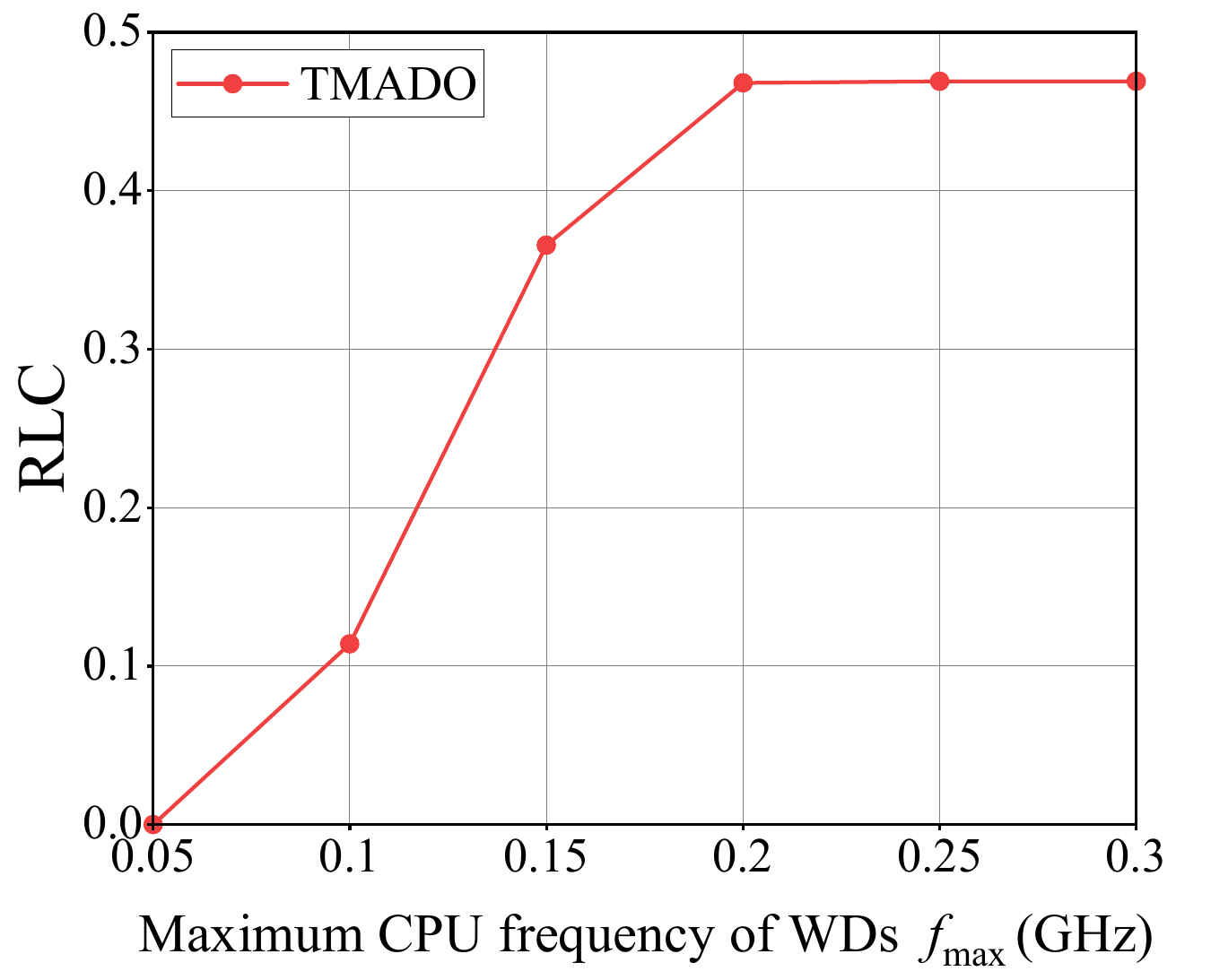}}
	\caption{(a) The EP by HAPs under six schemes versus the maximum CPU frequency of WDs $f_{\max}$. (b) The ratio of the number of WDs in local computing mode to the total number of WDs that process computation data (RLC) under the proposed TMADO scheme versus the maximum CPU frequency of WDs $f_{\max}$.}
	\label{Exp5CPU} 
\end{figure}

\begin{figure}[t]
	\centering
	\subfloat[]{
		\includegraphics[width=4.25cm]{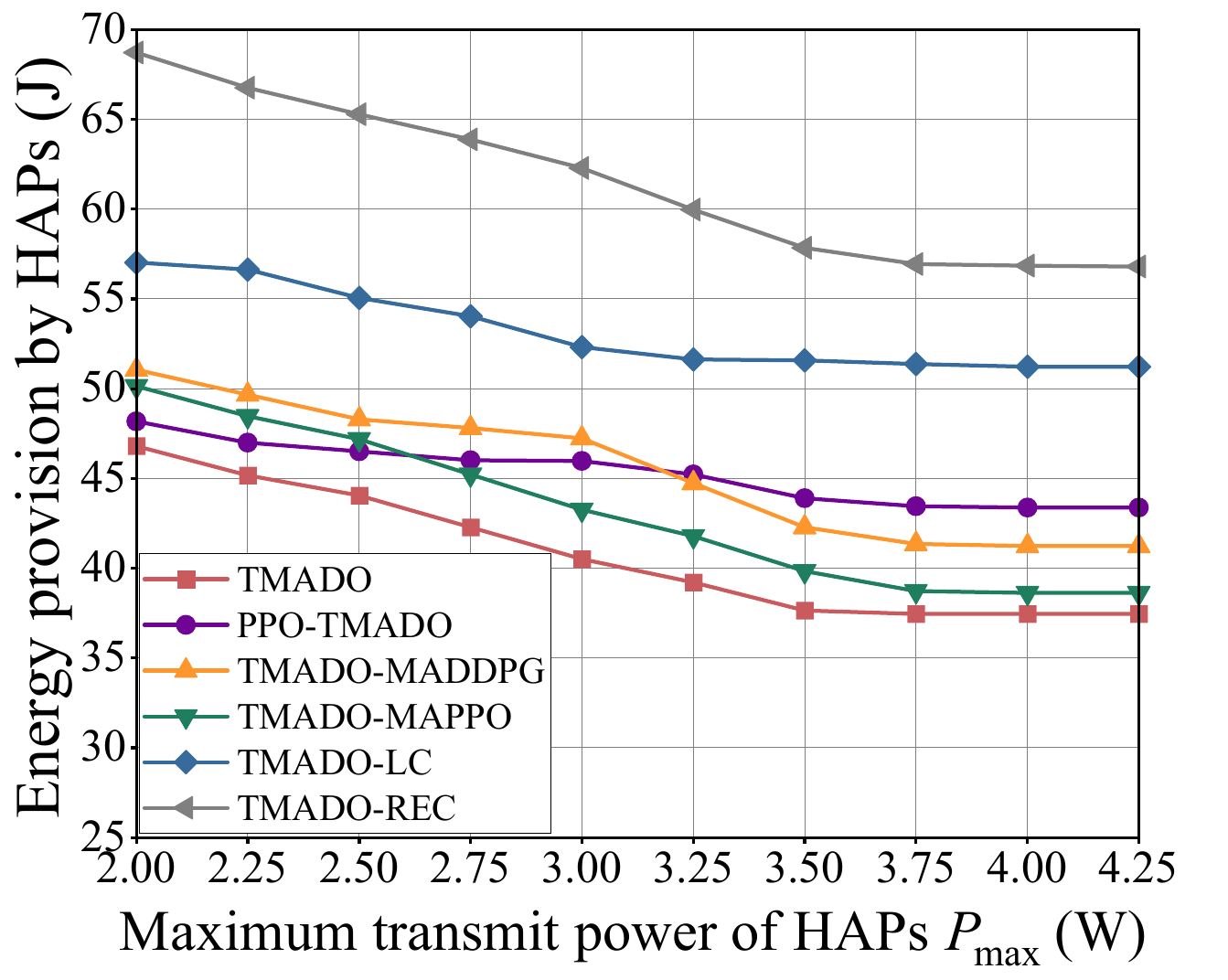}}
	\subfloat[]{
		\includegraphics[width=4.25cm]{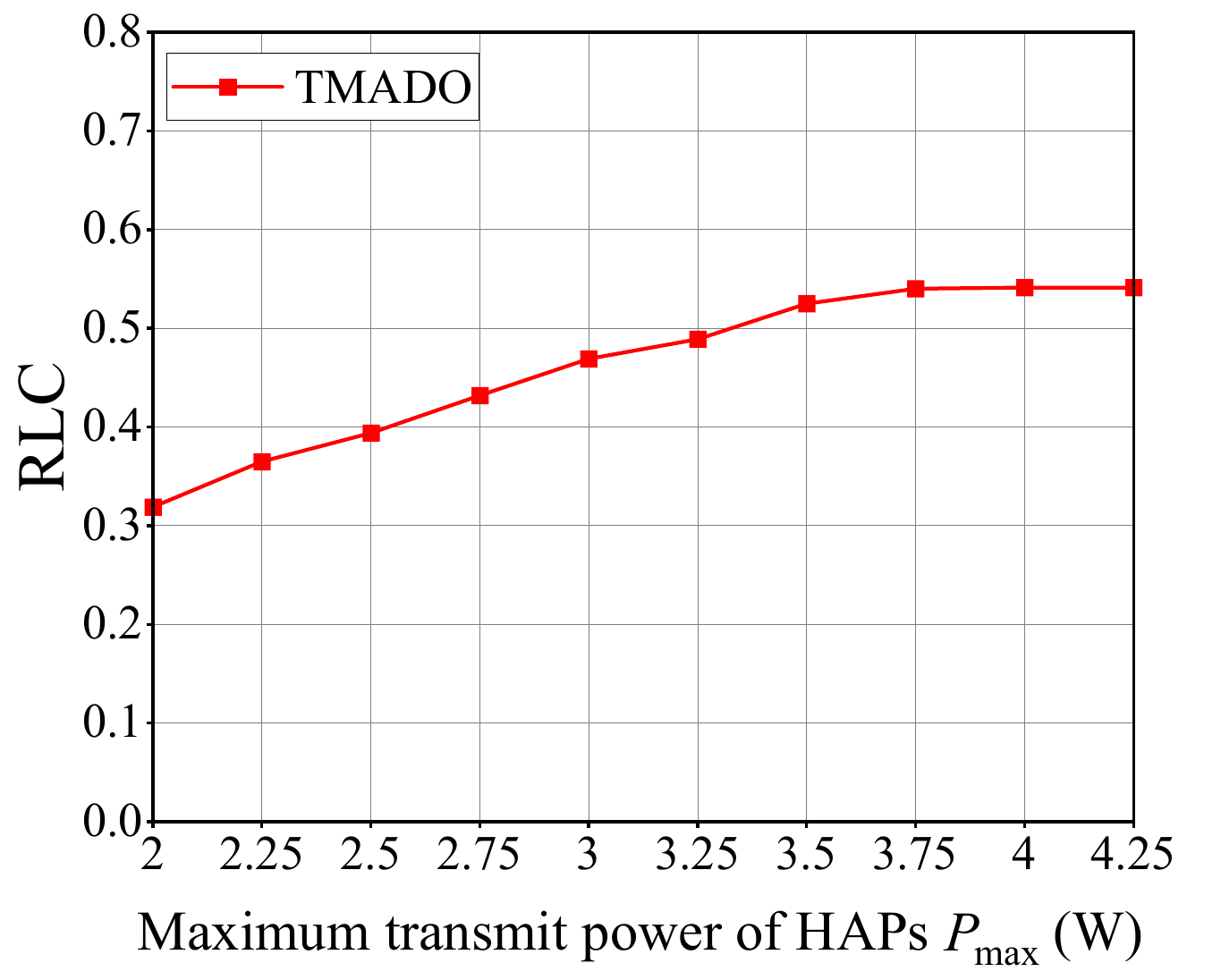}}
	\caption{(a) The EP by HAPs under six schemes versus the maximum transmit power of HAPs $P_{\max}$. (b) The RLC under the proposed TMADO scheme versus the maximum transmit power of HAPs $P_{\max}$.}
	\label{Exp8P} 
\end{figure}

In the following, we first show the convergence of the proposed TMADO scheme in Figs. \ref{Exp1high}-\ref{Exp1low} and then reveal the impacts of crucial parameters, such as the number of HAPs/WDs and the radius of the transmission zone,
on the energy provision by HAPs in Figs. \ref{Exp3Number}-\ref{Exp8P}.
To evaluate the proposed TMADO scheme in terms of the energy provision by HAPs, we provide five comparison schemes as follows.

\begin{itemize}
	\item Proximal policy optimization (PPO)-TMADO scheme:
	We use the PPO \cite{WYN2022JSAC} to obtain $\{ \boldsymbol{{P}_{{h}}(t)}, \alpha(t), \mathcal{N}_\text{-1}(t) \}$, and use the proposed TMADO
	scheme to obtain  $\{\boldsymbol{\mathcal{X}(t)}, \boldsymbol\tau_{o}(t), \boldsymbol{f(t)}\}$
	\item TMADO-MADDPG scheme:
	MADDPG is DDPG with the centralized state value function \cite{ZH2022HMAD}. We use the proposed TMADO scheme to obtain $\{ \boldsymbol{{P}_{{h}}(t)}, \alpha(t), \mathcal{N}_\text{-1}(t)\}$, and use the MADDPG to obtain $\{\boldsymbol{\mathcal{X}(t)}, \boldsymbol\tau_{o}(t), \boldsymbol{f(t)}\}$.
	\item TMADO-MAPPO scheme:
	MAPPO is PPO with the centralized state value function \cite{HZW2023TMC}. We use the proposed TMADO scheme to obtain $\{ \boldsymbol{{P}_{{h}}(t)}, \alpha(t), \mathcal{N}_\text{-1}(t)\}$, and use the MAPPO to obtain $\{\boldsymbol{\mathcal{X}(t)}, \boldsymbol\tau_{o}(t), \boldsymbol{f(t)}\}$.
	\item TMADO-random edge computing (REC) scheme: All WDs offload the computation data to random HAPs. We set the radius of the transmission zone $R_{t} =+\infty$ m. Besides, we use the proposed TMADO scheme to obtain $\{ \boldsymbol{{P}_{{h}}(t)}, \alpha(t), \mathcal{N}_\text{-1}(t)\}$.
	\item TMADO-local computing (LC) scheme: All WDs adopt local computing mode. Besides, we use the proposed TMADO scheme  to obtain $\{ \boldsymbol{{P}_{{h}}(t)}, \alpha(t), \mathcal{N}_\text{-1}(t)\}$.
\end{itemize}

{Fig. \ref{Exp0HyperPara} plots the energy provision (EP) versus $u$. The hyperparameter $u$ directly determines the reward of the WD that successfully processes the computation data. Since the number of the arrived data packets in WD$_n$, i.e., $D_n(t)$, follows an independent Poisson point process with rate $\lambda = 50$ in the simulation, the value of $u$ in (\ref{Psi2}) is reformulated as $M P_{\max} T + \mathop{max}\limits_{m}\{e_m\} D_p \lambda$, which is equal to $3.6 + 0.05 = 3.65$.
We observe that $u = 3.65$ reaches the minimum EP, which validates the rationality and superiority of the selected $u$. 
The reason is that, when $u$ is small, WDs tend to select the action of failing to process the computation data, and then HAPs need to provide more energy to satisfy the computation data demand.
When $u$ is large, WDs tend to select the action of successfully processing the computation data, but are hard to make the offloading decisions, due to the indistinguishableness in the reward of selecting edge computing mode and that of selecting local computing mode. Hence HAPs also need to provide more energy to satisfy the computation data demand.}

Fig. \ref{Exp1high} plots the EP versus training episodes with four values of the learning rate for the high-level agent. We observe that, the proposed TMADO
scheme with learning rate $2 \times 10^{-5}$ achieves smaller EP than that
with learning rates $2 \times 10^{-3}$ and $2 \times 10^{-4}$, and converges faster than that with learning rate $2 \times 10^{-6}$. Thus we adopt $2 \times 10^{-5}$ as the learning rate of the high-level agent.
Fig. \ref{Exp1low} plots the mean episode rewards of WDs versus training episodes. We observe that, the mean episode rewards of WDs converge after 450 training episodes. This observation verifies the advantage of the TMADO scheme for WDs to achieve a stable policy.

Fig. \ref{Exp3Number} plots EP under the proposed TMADO scheme versus the number of HAPs $M$ and that of WDs $N$.
We observe that, EP decreases with $N$. {This is due to the reason that, the increase of $N$ means that more WDs could harvest energy from HAPs, and accordingly more WDs have sufficient energy to process the computation data in local computing mode or edge computing mode. Thus, with fixed $D_{th}$, HAPs could provide less energy for WDs to satisfy computing delay and computation data demand constraints.}
 We also observe that, EP decreases with $M$. This is due to the reason that, the increase of $M$ reduces the average distance between HAPs and WDs, and increases the probability that WDs successfully process the computation data in edge computing mode. Thus HAPs could provide less energy for WDs to satisfy computing delay and computation data demand constraints.
 

 
 
Fig. \ref{Exp3Number-WD} and Fig. \ref{Exp3Number-HAP} plot EP under six schemes versus the number of WDs $N$ with the number of HAPs $M=6$, and EP versus the number of HAPs $M$ with the number of WDs $N=40$, respectively.
We observe that, the proposed TMADO scheme achieves the minimum EP compared with other comparison schemes, which validates the superiority of the proposed TMADO scheme in terms of EP.
This is due to the reason that, the DDPG takes advantage of learning continuous policies to solve the WCDO subproblem, while the IPPO employs the individual state value function of each WD to reduce the impact of interference from irrelevant state information of other WDs, which is better for solving the ODO subproblem than the algorithms with the centralized state value function of WDs, i.e., the MADDPG and MAPPO.
We also observe that, EP under the TMADO-REC scheme is larger than that under the TMADO-LC scheme for $N < 35$ in Fig.~\ref{Exp3Number-WD} and $M  < 5$ in Fig.~\ref{Exp3Number-HAP}, but is smaller than that under the TMADO-LC scheme for $ N\geq 35$ in Fig.~\ref{Exp3Number-WD} and $M \geq 5$ in Fig.~\ref{Exp3Number-HAP}.
This is due to the reason that, when $N$ or $M$ is small, the average distance between HAPs and WDs is large. Then WDs under the TMADO-REC scheme would need more energy to offload the computation data to random HAPs to satisfy computing delay and computation data demand constraints. While when $N$ or $M$ is large, the average distance between HAPs and WDs would be small, which increases the number of accessible HAPs to WDs. Accordingly, HAPs could provide less energy for WDs under the TMADO-REC scheme to satisfy computing delay and computation data demand constraints.

Fig. \ref{Exp4Bandwidth} plots EP versus the transmission bandwidth $B$. We observe that, EP under the proposed TMADO
scheme is the minimum, which validates the advantage of the proposed TMADO scheme in terms of EP.
We also observe that, EP under the TMADO-LC scheme remains unchanged with the transmission bandwidth,  and EP under the other schemes decreases with the transmission bandwidth.
This is due to the reason that, the increase of the transmission bandwidth shortens the duration that WDs offload the computation data to HAPs, and decreases the amount of the energy consumed by WDs in edge computing mode.
As edge computing mode is not considered in the TMADO-LC scheme, EP under the TMADO-LC scheme remains unchanged with the offloading bandwidth.

Fig. \ref{Exp7Demand} plots EP versus the computation data demand $D_{th}$.
We observe that, EP increases with $D_{th}$.
With higher value of $D_{th}$, WDs need to process larger amount of computation data in local computing mode or edge computing mode and make more offloading decisions, which requires more EP.

Fig. \ref{Exp6Possion} plots EP versus the radius of the transmission zone $R_{t}$.
We observe that, EP first decreases with $R_t$ and then increases with $R_{t}$.
{Apparently, the value of $R_t$ not only influences the number of accessible HAPs to WDs but also the observation space of low-level agents at WDs, and both of them increase with $R_t$. When $R_t$ is small, the accessible HAPs are not enough for WDs to find the optimal offloading decisions, and then the number of accessible HAPs to WDs is the key factor influencing the EP by HAPs. Hence the EP by HAPs first decreases with $R_t$. 
When $R_t$ is large, more HAPs far away from WDs are included in the transmission zones, which means that more redundant observation information is included in the observation space of WDs, i.e., $\boldsymbol{o}^{l}_{n}(t), n \in \mathcal{N}$ in Section IV-B. Although the accessible HAPs are enough for WDs to find the optimal offloading decisions,  the observation space of WDs, i.e., $\boldsymbol{o}^{l}_{n}(t), n \in \mathcal{N}$ in Section IV-B, also becomes large, which in turn causes that the low-level agents at WDs are difficult to find the optimal offloading decisions. Hence, the EP by HAPs then increases with $R_t$.}

Fig. \ref{Exp5CPU} (a) plots EP versus the maximum CPU frequency of WDs $f_{\max}$.
We observe that, EP under the TMADO-REC scheme remains unchanged, and EP under the other schemes decreases with the maximum CPU frequency of WDs for $f_{\max} \leq 0.2$ GHz, and remains unchanged with the maximum CPU frequency of WDs for $f_{\max} > 0.2$ GHz.
This is due to the reason that, WDs under the TMADO-REC scheme
are always in edge computing mode, and the energy consumption
is independent of the maximum CPU frequency of WDs. Besides,
when $f_{\max}$ is small, the computing delay constraint for WDs in local computing mode can not be satisfied, and $f_{\max}$ is the bottleneck limiting WDs to choose local computing mode. With the increase of $f_{\max}$, WDs have higher probabilities to choose local computing mode, i.e., the ratio of the number of WDs in local computing mode to the total number of WDs that process computation data (RLC) becomes larger, as shown in Fig.~\ref{Exp5CPU} (b).  Hence EP first decreases with $f_{\max}$. However, when $f_{\max}$ reaches a certain value such as $0.2$ GHz in Fig.~\ref{Exp5CPU} (a), the optimal $f_n^*(t)$ in (\ref{close2}) is in the range of $[0, f_{\max}]$, and the RLC is no longer affected by $f_{\max}$, as shown in Fig.~\ref{Exp5CPU} (b). Hence EP keeps unchanged.
We also observe that, EP with $f_{\max} < 0.15$ GHz under the TMADO-LC scheme is not shown in Fig. \ref{Exp5CPU} (a), due to the reason that WDs with small $f_{\max}$ can not satisfy the computing delay constraint and the computation data demand $D_{th}$.

Fig. \ref{Exp8P} (a) plots EP versus the maximum transmit power of HAPs $P_{\max}$.
We observe that, the proposed TMADO scheme achieves the minimum EP compared with comparison schemes.
We also observe that, EP first decreases and then keeps unchanged with the maximum transmit power of HAPs $P_{\max}$. The reason is that, when $P_{\max}$ is small, such as $P_{\max}\leq 3.75$ W in Fig.~\ref{Exp8P} (a), WDs are in the energy-deficit state. In such a context, it is best for HAPs to transmit RF signals with the maximum transmit power $P_{\max}$, so that WDs harvest more energy to have more choices, i.e., local computing mode or edge computing mode. {Actually, 
the WDs with more energy, such as the WDs close to HAPs (i.e., with high channel gains), prefer local computing so as to avoid the energy consumption by HAPs for processing the offloaded computation data. While the WDs with less energy, such as the WDs far away from HAPs (i.e., with low channel gains), can not support local computing and have to offload the computation data to HAPs. Then HAPs consume their energy to process the computation data. With the increase of $P_{\max}$, WDs harvest larger amount of energy. More WDs, especially the WDs with high channel gains, have sufficient energy to perform local computing, which is beneficial to reduce the EP by HAPs for processing the offloaded computation data. Hence the RLC increases with $P_{\max}$, as shown in Fig.~\ref{Exp8P} (b).}
Accordingly, EP first decreases with  $P_{\max}$. When  $P_{\max}$ reaches a certain value, i.e., $3.75$ W in Fig.~\ref{Exp8P} (a), WDs have two modes to process the computation data, and the optimal transmit power of HAPs is not the maximum. Hence EP keeps unchanged for $P_{\max}\geq 3.75$ W.

\section{Conclusion}
This paper studied the long-term energy provision minimization problem in a dynamic multi-HAP WP-MEC network under the binary offloading policy. We mathematically formulated the optimization problem by jointly optimizing the transmit power of HAPs, the duration of the WPT phase, the offloading decisions of WDs, the time allocation for offloading and the CPU frequency for local computing, subject to the energy, computing delay and computation data demand constraints. To efficiently address the formulated problem in a distributed way, we proposed a TMADO framework with which each WD could optimize its offloading decision, time allocation for offloading and CPU frequency for local computing. Simulation results showed that the proposed TMADO framework achieves a better performance in terms of the energy provision by comparing with other five comparison schemes. 
{This paper investigated the fundamental scenario with single-antenna HAPs. As one of the future research directions, we will further study the scenario with multi-antenna HAPs by jointly considering the beamforming technique and channel allocation. 
}

\vfill
\end{document}